\documentclass{aa}

\usepackage{epsfig,graphicx,natbib}
\usepackage{longtable}
\usepackage{amssymb}
\usepackage{amsmath}
\usepackage{mathrsfs} 
\usepackage{color}
\usepackage{subcaption}
\usepackage{booktabs}
\usepackage{xspace}
\usepackage{float}
\usepackage{longtable}
\usepackage{breqn}
\usepackage{algpseudocode}
\usepackage{algorithm}
\usepackage{hyperref}
\usepackage[raggedright]{sidecap}

\usepackage{todonotes}

\newcommand{\RR}{\mathbb{R}}

\newcommand{\sgra}{Sgr~A$^*$\xspace}
\newcommand{\mes}{M87$^*$\xspace}

\renewcommand{\textbf}[1]{#1}
\renewcommand{\textit}[1]{#1}

\defcitealias{Issaoun2022}{Issaoun, Wielgus et al. 2022}
\defcitealias{Jorstad2022}{Jorstad, Wielgus et al. 2022}
\defcitealias{Mueller2023}{Paper I}
\defcitealias{Mus2023}{Mus (PhD Thesis)}
\defcitealias{Mus2023b}{Paper II}
\defcitealias{Zhang2008}{Zhang \& Li (2007)}
\defcitealias{Pardalos2017}{Pardalos et al. 2017}

\begin{document}

\title{Multiobjective optimization for scattering mitigation and scattering screen reconstruction in very long baseline interferometry observations of the Galactic Center}

\author{Alejandro Mus \inst{1,2,3,4,5}
\and Teresa Toscano \inst{6}
\and Hendrik M\"uller \inst{7,8}
\and Guang-Yao Zhao\inst{8,6}
\and Andrei Lobanov \inst{8}
\and Ciriaco Goddi \inst{1,9}
}

\institute{
  Dipartimento di Fisica, Università degli Studi di Cagliari, SP Monserrato-Sestu km 0.7, I-09042 Monserrato, Italy\email{alejandro.musmejias@unica.it}
  \and INAF-Istituto di Radioastronomia, Via P. Gobetti 101, I-40129 Bologna, Italy
  \and SCOPIA Research Group, University of the Balearic Islands, Dpt. of Mathematics and Computer Science, Crta. Valldemossa, Km 7.5, Palma, E-07122, Spain
  \and Health Research Institute of the Balearic Islands (IdISBa), Palma, E-07122, Spain
  \and Artificial Intelligence Research Institute of the Balearic Islands (IAIB), Palma, E-07122, Spain
  \and Instituto de Astrof\'{i}sica de Andaluc\'{i}a CSIC, Apartado 3004, 18080 Granada, Spain
  \and Jansky Fellow of National Radio Astronomy Observatory, 1011 Lopezville Rd, Socorro, NM 87801, USA
  \and Max-Planck-Institut für Radioastronomie, Auf dem Hügel 69, D-53121 Bonn (Endenich), Germany
 \and 
  Instituto de Astronomia, Geof\'isica e Ci\^encias Atmosf\'ericas, Universidade de S\~ao Paulo, R. do Matão, 1226, S\~ao Paulo, SP 05508-090, Brazil
}

\date {Received  / Accepted}

\authorrunning{Mus, A.}
\titlerunning{Multiobjective optimization for scattering mitigation in very long baseline interferometry}

\abstract
%context heading (optional)
{Imaging reconstruction of interferometric data is a hard and ill-posed inverse problem. Its difficulty is increased for observations of the Galactic Center, which is obscured by a scattering screen. This is because the scattering breaks the one-to-one correspondence between images and visibilities.} 
% aims heading (mandatory)
{Solving the scattering problem is one of the greatest challenges in radio imaging of the Galactic Center. We present a novel strategy for mitigating its effect and for constraining the screen itself using multiobjective optimization.}
% methods heading (mandatory)
{We exploited the potential of evolutionary algorithms to describe the optimization landscape with the aim to recover the intrinsic source structure and the scattering screen that affects the data.}
% results heading (mandatory)
{We successfully recovered the screen and the source in a wide range of simulated cases, including the speed of a moving screen at 230 GHz. Particularly, we recovered a ring structure in scattered data at 86\,GHz.}
% conclusions heading (optional)
{Our analysis demonstrates the huge potential that recent advancements in imaging and optimization algorithms \textbf{offer} in recovering image structures, even in weakly constrained, degenerated, and possibly multimodal settings. The successful reconstruction of the scattering screen opens the window to event-horizon scale works on the Galactic Center \textbf{at 86\,GHz up to 116\,GHz} and to the study of the scattering screen itself.}

\maketitle

\section{Introduction}\label{sec:intro}
Supermassive black holes (SMBH) are found or are expected to be found in the centers of most galaxies~\citep{Richstone1998}. The accretion of matter onto the central object causes the regularly observed periods of activity that significantly impact the intergalactic and circumgalactic environments and the evolution of their host galaxy. Additionally, SMBHs are used as a laboratory to study the physics of gravity itself. 
In this context, a primary laboratory is the SMBH in our own Galaxy, Sagittarius~A*, or \sgra \citep{Goddi2017}.

\sgra is one of the key targets of 
the Event Horizon Telescope (EHT), a worldwide network of radiotelescopes  that employs very long baseline interferometry (VLBI) at millimeter and submillimeter wavelengths \citep{eht2019b,Raymond2024}. The unprecedented angular resolution achieved by the EHT enabled us to capture the first images of SMBHs, that is, the SMBH at the center of the giant elliptical galaxy M87 \citep{eht2019a, eht2019b, eht2019c, eht2019d, eht2019e, eht2019f, eht2023} and the SMBH at the center of our own galaxy, in total intensity~\citep{eht2022a, eht2022b, eht2022c, eht2022d, eht2022e, eht2022f} and in polarized light~\citep{eht2021a, eht2021b, eht2024b, eht2024c}. In both targets,  EHT observations revealed a ring-like feature, which was interpreted as the theoretically predicted signature of the black hole shadow.
In the case of \mes, a similar ring-like feature has subsequently also been detected with the Global Millimeter VLBI Array (GMVA) at 86 GHz \citep{Lu2023, Kim2024}. \textbf{Multifrequency} observations of the innermost regions surrounding the central SMBH are crucial for characterizing the physical conditions of the  accreting plasma and for constraining models for black hole accretion  \citep{Moscibrodzka2014,Moscibrodzka2016}. 

However, \textbf{a multifrequency} analysis like this remains \textbf{limited} for \sgra as a result of an additional effect: interstellar scattering. Interstellar scattering is caused by variations in the electron density along the line of sight to the target. It results in multipath propagation, which broadens radio images and pulse profiles of objects that are located in or behind the Galactic plane. In certain directions, the scattering is significantly stronger than what is predicted by large-scale Galactic electron distribution models~\citep{Taylor1993, Cordes2002}. This pronounced scattering has long been linked to Galactic \textsc{Hii} regions and supernova remnants ~\citep{Litvak1971,Little1973}. Being situated at the center of our Galaxy, \sgra is chiefly affected by intense scattering. The size of its radio image increases with $\lambda^2$~\citep[e.g.][]{Bower2004, Issaoun2019}, consistent with predictions for a thin scattering screen~\citep{Blandford1985}.

The interferometric phase information is strongly affected by the presence of scattering. As a consequence, scattering-induced distortions on the observed data can significantly degrade the quality of the images, particularly at frequencies below \textbf{$86\,$GHz.} 
In this line, \textbf{several strategies} have been developed to mitigate the scattering effects and improve the quality and robustness of the image~\citep{Fish2014, Johnson2016, Kouroshnia2025}. These algorithms characterize the scattering kernel and incorporate a functional within the imaging deconvolution process to correct for the associated distortions. First,~\cite{Fish2014} proposed to apply a Wiener filter~\citep{Wiener1949} to the scattering kernel to increase the \textbf{signal-to-noise ratio} (SNR). In contrast,~\cite{Johnson2016} proposed to find the source and screen discrete reconstructions that fit the data best. As a byproduct, this algorithm is able to recover an image of the scattering-corrected observed source and an approximation of the scattering screen. The latter  holds significant astrophysical information on its own and supports efforts to detect transient phenomena 
toward the Galactic Center \citep{Caleb2022, Beniamini2023}.
One of the main challenges in this context is the potential degeneracy between the scattering screen and the image.
This led to the development of imaging algorithms based on multimodal image posteriors. 
In particular, we  base our analysis on the multiobjective evolution proposed by \citet{Mueller2023c} and then was extended by \citet{Mus2024b, Mus2024c}.

We present in this work the design, implementation, and testing of novel algorithms to mitigate the impact of (potentially dynamic) scattering screens that affects VLBI observations by  
recovering the \textbf{intrinsic black hole structure images} at lower frequencies and the screen itself. 

\textbf{Compared to traditional noise-inflation and deblurring techniques, such as those used by~\cite{eht2022c, eht2024b} or even standard unimodal exploration~\cite{Johnson2016}, our novel strategy effectively explores the family of local solutions. This is an essential feature for addressing the highly nonlinear and ill-posed problem of VLBI imaging and screen modeling. Unlike methods that are constrained to a specific screen model (and thus, to a fixed intrinsic source structure), the flexibility of our approach provides a mathematically rigorous solution to the modeling and mitigation problem.}

\textbf{This paper is structured as follows. } We present the background theory on VLBI imaging and scattering theory in Sec. \ref{sec2}. In Sec. \ref{sec3} we discuss the problem modeling and imaging algorithm, and we verify the reconstruction with a static and dynamic screen and a static source in Sec. \ref{sec4}. In Sec. \ref{sec:86ghz} we discuss the potential of recovering the scattering \textbf{screen} and the black hole shadow after proper descattering at 86 GHz. We show the marginal distributions of the scattering screen regularizers in Sec. \ref{sec:marginal}, and we finally discuss future prospects in Sec. \ref{sec:future}.

\section{VLBI and imaging background} \label{sec2}
\subsection{VLBI measurements}
In a VLBI array, multiple antennas observe the same source at the same time, and the recorded signals are then correlated. These correlated signals from the antennas (visibilities) are approximately the Fourier transform of the true sky brightness distribution at a spatial frequency determined by the baseline separating the antennas in an antenna-pair projected on the sky-plane. 
In particular, the visibility function $\mathcal{V}(u,v)$, representing the correlated signal between an antenna pair, is governed by the van Cittert-Zernike theorem~\citep{Thompson2017},
\begin{align}
\mathcal{V} (u, v) = \int \int I(l, m) e^{-2 \pi i (l u + m v)} dl dm. \label{eq: vis}
\end{align}
This equation expresses that the true sky brightness distribution, $I(l, m)$, and the observed visibilities are related as a Fourier pair. Here, $(u, v)$ coordinates are defined by the relative baselines between antennas in the plane of the sky and $\left(l,m\right)$ direction cosines
measured with respect to the $(u, v)$ axes. However, it is crucial to recognize that in practical scenarios, the Fourier domain is not fully sampled, leading to sparse coverage in the $(u, v)$-plane, particularly in VLBI contexts. This incomplete sampling makes the process of reconstructing the sky brightness distribution from the observed visibilities an ill-posed inverse problem. Additionally, these visibilities are subject to calibration errors and thermal noise. To mitigate these challenges, it is essential to account for calibration effects. Direction-independent calibration errors are modeled through station-based gain factors, $g_i$. The observed visibilities at a given time $t$ for an antenna pair $i,j$ are thus expressed as
\begin{align}
V(i,j,t) = g_{i,t} g_{j,t}^{*} \mathcal{V}(i,j,t) + N\left(i,j,t\right),
\end{align}
where $g_i$ and $g_j$ are the station-based gains, and $N\left(i,j,t\right)$ represents the random noise associated with the baseline.

These difficulties are exasperated in  sparse VLBI networks such as the EHT, which additionally operates at high radio frequencies, rendering the data calibration and imaging processes particularly challenging. 
All these difficulties have required the 
development of a wide range of novel algorithms for image \textbf{reconstruction} that we describe in detail in  the next subsection.

\subsection{Outline of the imaging}

 The imaging  algorithms employed in the EHT can be roughly categorized into five groups: techniques based on Bayesian exploration \citep{Broderick2020b, Tiede2022}, the regularized maximum likelihood technique (RML) \citep{Akiyama2017, Akiyama2017b, Chael2018}, global search techniques by multiobjective evolution \citep{Mueller2023c, Mus2024c}, compressive sensing \citep{Mueller2022, Mueller2023b}, and inverse modeling with novel \textbf{developments, for instance, by \citet{Mueller2023a}}. The majority of these algorithms \textbf{achieves} a moderate degree of super-resolution, that is, they robustly recover structures smaller than the original resolution limit of CLEAN~\citep{Hogbom1974,Clark1980,Cornwell2008}. This has proven to be crucial for a reliable reconstruction of structures at the spatial scales of the event horizon. 

The RML method is one of the most frequently used imaging techniques in the EHT \citep{eht2019d, eht2021a, eht2022c, eht2023, eht2024a, eht2024b}. Multiple data-fidelity terms and regularization terms are balanced against each other by a relative weighting. This weighted sum is then minimized. 
The solution of this optimization problem is the model that describes the data better.  The data term takes into account how well one solution is described by the data. Frequently used data terms include $\chi^2$ metrics for observed data products, for example, the visibilities or closure products. The regularization terms in turn describe the plausibility of the recovered solution. Standard choices include the total variation (promoting piecewise smooth functions), the total squared variation (promoting smoothness), the l1-norm (promoting sparsity), the l2-norm (reducing the overfitting), or an entropy functional.

It is crucial to use the correct relative weighting of the data and regularization terms for RML methods. The EHT has applied a strategy of thorough surveying a \textbf{sufficient subset of possible parameter combinations}. Multiobjective optimization aims at presenting a different strategy. The concept of optimality of a weighted sum of objectives is replaced by the concept of Pareto optimality in a multiobjective problem formulation: A solution is called Pareto optimal when the further optimization of one objective (e.g., the data term) automatically has to worsen the scoring in another objective (e.g., in a regularization term). The set of all Pareto-optimal solutions is referred to as the Pareto front. 

The algorithm MOEA/D\footnote{Multiobjective Evolutionary Algorithm Based on Decomposition.} is based on the genetic evolution that approximates the Pareto front \citep{Zhang2008, Li2009}. It has been adapted for VLBI objectives by \citet{Mueller2023c}, who also showed that the Pareto front in VLBI experiments has a special structure: It is separated into multiple clusters, which are interpreted as locally optimal modes of the potentially multimodal imaging problem. Furthermore, the cluster with the best solution has in practice been found to be the cluster that lies closest to the ideal point \citep[see ][for more details on this construction]{Mueller2023c}. 
This algorithm is in active use for the imaging in extremely weak constrained settings or in problems in which internal degeneracy may cause a multimodal posterior. This includes the reconstruction of the polarization structure from the closure traces \citep{Mueller2024b} or the imaging of solar flares with the Spectrometer/Telescope for Imaging X-rays (STIX) instrument \citep{Mueller2024a}.

The variant MO-PSO\footnote{standing from Multiobjective Optimization and Particle Swarm Optimization.} is much faster and more accurate than MOEA/D that was proposed by \citet{Mus2024c}. Instead of reconstructing the full image structure by an evolutionary algorithm, the code navigates the Pareto front by an evolutionary algorithm (particle swarm optimization) and solves the scaled problems by fast L-BFGS-B (limited memory Broyden–Fletcher–Goldfarb–Shanno) minimization. The Pareto front is searched in order to minimize the distance of a Pareto-optimal solution to the ideal point. Thus, instead of recovering the full Pareto front, MO-PSO directly provides its final best image reconstruction, for which it spends less time and fewer computational resources.

\subsection{Scattering}

In this section, we give an overview of the theory of and motivation for the thin-screen scattering. We focus on the relevant aspects related to the goals of our strategy. For further details, we refer to the reviews by~\cite[][]{Rickett1990, Narayan1992, Johnson2015, Thompson2017}.

When a source is observed through a medium with spatial variations in its refractive index, it becomes distorted. Refractive inhomogeneities cause different \textbf{regions of an image} associated with the source to be steered and focused differently while surface brightness is maintained~\citep{Born1980}. In particular, radio-wave scattering in the ionized interstellar medium (ISM) is due to density inhomogeneities because the refractivity in a plasma is approximately proportional to the local electron density~\citep{Jackson1999}.

In general, scattering is often described as resulting from turbulent media localized in a single thin screen between the source and observer~\citep[see][for instance]{Bower2014,Dexter2017,Psaltis2018}. We work throughout under the assumptions of one unique thin screen that can be approximated by $n$ layers, although the \textbf{number of screens} in the line of sight is still an opened question~\cite[e.g.][]{Dexter2017}.

When $r$ is a two-dimensional transverse coordinate on the screen, this screen introduces a stochastic position-dependent phase shift $\phi\left(r\right)$ of the incoming radiation, without altering the amplitude of the waves. If this screen follows \textbf{a homogeneous} Gaussian distribution, the scattering can be quantified in two complementary ways: by the structure function of the phase fluctuations, \( D_\phi(r) = \langle [\phi(r + r') - \phi(r')]^2 \rangle \), or by the power spectrum of the phase fluctuations, \( P_\phi(q) \propto |q|^{-(\alpha + 2)} \). We assumed $\alpha$ to be 1.38\footnote{We used the default value, 1.38, of \texttt{StochasticOptics}~\citep{Johnson2016}.}. 
These approaches are related by a Fourier transform, \( P_\phi(q) \propto q^2 D_\phi(q) \), with a prefactor that makes \( P_\phi(q) \) dimensionless and independent of the observing wavelength (\(\lambda\)).

Diffractive and refractive scattering effects on images can be treated as distinct phenomena~\citep{Blandford1985}. Diffractive effects dominate at small scales (approximately \( r_0 \)) and \textbf{are well approximated by their ensemble average, that is, the expectation value of the measured complex visibilities when averaging over many realizations of the scattered phases~\citep{NarayanGoodman1989, GoodmanNarayan1989}. This averaging effectively blurs the intrinsic image with a kernel \( G(r) \) (the seeing disk).} This kernel is most naturally expressed in the visibility domain as \( \tilde{G}(\mathbf{b}) = e^{-\frac{1}{2} D_\phi(\mathbf{b})} \), where \(\mathbf{b}\) corresponds to the physical length of an interferometric baseline. Refractive effects can be modeled using a geometrical optics framework, where gradients of the large-scale refractive modes of the phase screen steer and focus the ensemble-average image.

The single-epoch scattered image \( I_a(r) \) is related to the unscattered image \( I_{\text{src}}(r) \) via \cite{JohnsonNarayan2016, Johnson2016}
\[ 
I_a(r) = I_{\text{src}}(r) * G(r) + (\nabla \phi_r(r)) \cdot [I_{\text{src}}(r) * \nabla G(r)]. 
\]
In these expressions, \( \nabla \) represents the two-dimensional transverse gradient on the phase screen. The term \( I_{\text{src}}(r) * G(r) \), where the asterisk denotes spatial convolution, represents the blurring effect due to scattering and is referred to as the ensemble-average image, \( I_{\text{ea}}(r) = I_{\text{src}}(r) * G(r) \). The second term, \( (\nabla \varphi_r(r)) \cdot [I_{\text{src}}(r) * \nabla G(r)] \), accounts for additional distortions introduced by variations in the phase screen. This term captures the influence of the turbulent phase gradient across the scattering screen, further modulating the observed image through spatially dependent distortions.

For each image, \( r \) is a transverse coordinate at the distance of the scattering screen \( D \) (not the distance of the source \( D + R \)), so that the corresponding angular scales are \( \theta = \frac{r}{D} \). For simplicity, we use \( \phi(r) \) throughout to denote the refractive phase screen \(\phi_r(r)\). We denote the forward operator applying the scattering with $\Phi$ for the remainder of this paper, that is,
\begin{align}
    \Phi: \left[ I_{\text{src}}, \phi \right] \mapsto I_{a}.
\end{align}

The Strehl ratio, $0<S\leq1$, is a commonly used metric in the optical literature to quantify image degradation due to scattering. It is defined as the ratio of the peak intensity of the measured point-spread function (PSF), which includes the effects of scattering, to the peak intensity of the ideal diffraction-limited PSF, which assumes no scattering.

This concept was extended to radio interferometry by~\citep{Johnson2016}, and it is defined as
\[
S \simeq \dfrac{\theta_{uv}}{\sqrt{\theta_{uv}^{2} + \theta_{scatt}^2}},
\]
where $\theta_{uv}$ is the full width at half maximum (FWHM) of the beam of the array, and $\theta_{scatt}$ is the FWHM of $I_a$. \cite{Johnson2016} Fig. 1 shows the expected $S$ for the different arrays.

\section{Problem statement}\label{sec3}

In this section, we introduce the mathematical optimization formulation we used to model the imaging problem, along with the strategy we employed. We adopted a multiobjective optimization approach as presented by~\cite{Mus2024c}. For the sake of simplicity, we refer to the aforementioned paper for the mathematical formulation of the regularizers.

\subsection{Modeling}

Let $F\left(x\right):=\left(f_1\left(x\right),\ldots,f_n\left(x\right)\right)$ be an $n$-vector of scalar objective functions $f_i:\RR^p\to\RR$, and let $\mathcal{X}\subseteq\RR^{p}_{+\infty}$ denote the feasible set of decision vectors $x$. We then consider the multiobjective optimization problem

\begin{problem}[MOP]
  \begin{equation*}
  \label{prob:mop_ours}
  \tag{$\text{MOP}$}
    \begin{aligned}
      & {\text{min}}
      & & F\left(x\right):=\left(f_1\left(x\right),\ldots,f_n\left(x\right)\right),\\
      & \text{subject to}
      & & x \in\mathcal{X} \subseteq \mathbb{R}^{p}_{+\infty}.\\
    \end{aligned}
  \end{equation*}
\end{problem}
Following the strategy presented by~\cite{Mueller2023c, Mus2024b, Mus2024c}, we solved Prob.~\eqref{prob:mop_ours} by scalarization, that is, by rewriting the problem as the unconstrained minimization problem,

\begin{problem}[MOP scalarized]
  \begin{equation*}
  \label{prob:mop_ours_scalar}
  \tag{$\text{MOP Scalar}$}
    \begin{aligned}
      & \underset{x \in \mathcal{X} \subseteq \mathbb{R}^{p}_{+\infty}}{\text{min}}
      & & \tilde{F}\left(x\right):=\displaystyle{\sum_{i}^{n}}\alpha_i f_i\left(x\right),\quad \alpha_i\geq0,\forall i=1,\ldots,n.\\
    \end{aligned}
  \end{equation*}
\end{problem}

We recovered the images by using a wavelet dictionary $\Psi$, that is, we modeled the total intensity by $I = \Psi \omega_I$\footnote{Hereafter, we assume $x=I$ to remark that $x$ represents an image.} (see Sec. \ref{sec:pipeline} for more details). For the dynamic reconstruction of a movie in full polarization and a movie of the scattering screen, we therefore recovered the following quantities at every frame:
\begin{enumerate}
    \item The wavelet coefficients describing the image in total intensity $\omega_I$. They are related to the image at every frame by the relation $I = \Psi \omega_I$.
    \item The phase screen $\phi$.
\end{enumerate}
A movie is represented by a sequence of single individual images. In Prob.~\eqref{prob:mop_ours} $x$ is therefore the vector,
\begin{align}
 x = \left[ \omega_I, \phi^1, \phi^2, \phi^3, ... \right],
\end{align}
where the superscript denotes the index of the frame in the movie.

The individual functionals are
\begin{align}
    & f_1 :=  R_{l1}\left(\Psi \omega_I\right),\label{eq:l1}\\
    & f_2 :=  R_{tv}\left(\Psi \omega_I\right),\\
    & f_3 :=  R_{tsv}\left(\Psi \omega_I\right),\\
    & f_4 :=  R_{l2}\left(\Psi \omega_I\right),\\
    & f_5 :=  R_{flux}\left(\Psi \omega_I\right),\\
    & f_6 :=  R_{entr}\left(\Psi \omega_I\right),\label{eq:simple}\\
    & f_{7} := \sum_{j \in frames} R_{SO}\left(\phi^j \right) \label{eq:SO},\\
    & f_{8} :=
    \begin{aligned}
    \sum_{j \in frames} S_{vis}\left(\Phi (\Psi \omega_I, \phi^j)\right) + S_{amp}\left(\Phi (\Psi \omega_I, \phi^j)\right) \\
    + S_{clp}\left(\Phi (\Psi \omega_I, \phi^j)\right) + S_{cla}\left(\Phi (\Psi \omega_I, \phi^j)\right).
    \end{aligned} \label{eq:dataterms}
\end{align}

Here, $f_1$ to $f_6$ are the standard static imaging regularizers. For their exact form, we \textbf{refer to \citet{Mueller2023c}}, 
\textbf{$f_{7}$} is the stochastic optics functional~\citep{Johnson2016}, 
and $f_8$ is the data functional. These are $\chi^2$ of the visibilities, amplitudes, closure phases, and closure amplitudes in total intensity. The data-fidelity terms are evaluated on the scattered movies, while the single regularizers are evaluated on the unscattered guess solutions.

By solving Prob.~\eqref{prob:mop_ours} via Prob.~\eqref{prob:mop_ours_scalar}, we obtain the (static/dynamic) reconstruction that along with its associated power spectrum provides the best fit to the data. In Sect.~\ref{subsec:screen_discretization} we model the power spectrum in detail.

\subsection{Pipeline}\label{sec:pipeline}

The pipeline we used for the reconstruction employs nature-based optimization strategies: MOEA/D~\citep{Mueller2023c, Mus2024b} and MO-PSO~\citep{Mus2024c}. The core concept of these strategies involves scalarizing the objective vector $F$ and iteratively solving a series of subproblems. The two approaches offer significant advantages, which we discussed throughout. For a more in-depth understanding, we refer to the cited references.

\subsubsection{MOEA/D}

To explore the Pareto-front reconstructions, the interferometric reconstruction problem is modeled by
\[
    \tilde{f_i} = f_i + f_{8},
\]
where $f_i$ are those defined from Eq.~\eqref{eq:l1} to Eq.~\eqref{eq:SO}, and $f_{8}$ is the corresponding data term functional associated with the problem.

With this formulation, every functional $\tilde{f_i}$ gives a set of solutions that fit the data and are only affected by the regularizer $f_i$.

To solve the full MO, we applied a genetic algorithm. That is to say, we first gridded the multidimensional space of functions and drew a random sample of individuals (in our case, images or movies). Then, we applied a mutation and cross-over operation over the individuals producing new populations~\citep{Mueller2023c}.

\subsubsection{MO-PSO}
For every subproblem, we applied MO-PSO~\citep[for instance][]{Du2016}, considering a swarm composed by the hyperparameters $\alpha_i$ that appear in Prob.~\eqref{prob:mop_ours_scalar} and that act on every regularizer. In this way, we let the swarm converge to their optimal position (equivalently, to the optimal hyperparameter combination), and then we solved the imaging RML problem with the optimal weights.

In contrast to MOEA/D, the MO-PSO formulation allowed us to obtain the marginal contribution of every regularizer. In other words, instead of exploring the whole effects of each regularizer over the data terms, it seeks the best compromise among all the functionals.

By letting the population evolve long enough, individuals converge to a set of nondominate solutions~\citep[see][for definition]{Pardalos2017, Mueller2023c} that approximates the Pareto front. Therefore, an explicit dependence (manifold) between data terms and regularizers can be obtained.

\section{Reconstruction of the scattering screen}\label{sec4}

In this section, we validate the algorithm performance at high \textbf{frequencies}, in particular, at 230\,GHz. In this regime, scattering corruptions are minimal, thereby providing a clear framework for assessing the efficacy of the mitigation strategy. In the subsequent sections~\ref{sec:86ghz} and Appendix~\ref{app:scatt_freq}, we investigate the algorithm behavior at lower frequencies, where the corruptions are stronger and the problem is more constrained.

\subsection{Discretizsation of the stochastic optics scattering screen}\label{subsec:screen_discretization}

First, we present the ~\citep{Johnson2016} theoretical framework on stochastic optics, followed by the strategy we used to mitigate the scattering.

\textbf{The \cite{Johnson2016} framework} represents \( \phi(r) \) in the Fourier domain, with uncorrelated complex Gaussian variable components
\[
\tilde{\phi}(\mathbf{q}) = \int \phi(\mathbf{r}) e^{-i 2 \pi \mathbf{q} \cdot \mathbf{r}} d\mathbf{r}.
\]
The time-averaged power spectrum is given by
\[
\langle |\tilde{\phi}(\mathbf{q})|^2 \rangle = A_{\phi} Q(\mathbf{q}),
\]
where \( A_{\phi} \) is the screen area over which the Fourier transform is computed~\citep[see, e.g.][]{Blandford1985}.

The power spectrum \textbf{associated with} the observed visibilities is given by 
\begin{equation}\label{eq:theo_ps}
\begin{aligned}
Q(\mathbf{q}) &= 2^\alpha \pi \alpha \frac{\Gamma(1 + \alpha/2)}{\Gamma(1 - \alpha/2)} \lambda^{-2} (r_{0,x} r_{0,y})^{-\alpha/2} \\
&\times 
\left[ \left( \frac{r_{0,x}}{r_{0,y}} \right) q_x^2 + \left( \frac{r_{0,y}}{r_{0,x}} \right) q_y^2 \right]^{-(1+\alpha/2)},
\end{aligned}
\end{equation}

where $x,y$ are the coordinates aligned with the major and minor axes of the diffractive kernel. We considered the screen to be located at $M=0.43$, as stated by~\cite{Bower2014}.

The scattering phase screen (in radians) was parameterized using an \( N \times N \) grid of Fourier coefficients \(\tilde{\phi}_{o,s}\),
\begin{equation*}
\phi_{l,m} = \dfrac{1}{\sqrt{A_{\phi}}}\sum_{o,s=0}^{N-1}\sqrt{Q\left(o,s\right)}\ \epsilon_{o,s}e^{i 2 \pi \left(lo + ms\right)/N},
\end{equation*}
where $\epsilon_{o,s} = \dfrac{\tilde{\phi}_{o,s}}{\sqrt{Q\left(o,s\right)}}$. In this way,~\cite{Johnson2016} parameterized the unknown phase screen as a set of $\frac{N^2-1}{2}$ variables.

After \( \epsilon_{o,s} \) was specified for a particular field of view, the corresponding scattering was computed for any desired observing wavelength. 
We adopted this framework to model Eq.~\eqref{eq:SO}.

Therefore, the mitigation can be summarized as follows: For the observation data, we tried to find the best image that fits it and whose power spectrum is close to Eq.~\eqref{eq:theo_ps}. For every reconstruction, we computed its power spectrum and minimized along the $\chi^2$ axis. We denote this functional as $R_{SO}$. Mathematically, 
\begin{equation*}
    R_{SO} := \sum\lvert \epsilon_{o,s}\rvert^ 2.
\end{equation*}

\cite{Johnson2017} solved the optimization problem
\begin{problem}[\cite{Johnson2017} stochastic optics version]
  \begin{equation*}
  \label{prob:Jso}
  \tag{$\text{JohnsonSO}$}
    \begin{aligned}
      & \underset{x\in D}{\text{min}}
      & &  R_{entr} - \alpha_1 f_{8} - \alpha_2 R_{SO},\\
    \end{aligned}
  \end{equation*}
\end{problem}

where $R_{entr}, f_{9}$ are the functionals described in Eq.~\eqref{eq:simple} and Eq.~\eqref{eq:dataterms}, respectively, and only one frame ($j=1$) was fixed and was applied without wavelet coefficients.\footnote{I~\cite{Johnson2016} only considered the visibilities as data terms, but the \texttt{\textbf{eht-imaging}} library ~\citep{Chael2016, Chael2018} has the extended version.}. \textbf{There are some limitations, as listed below.}
\begin{itemize}
    \item \textbf{The lack of multimodal exploration in the context of highly nonlinear nonconvex functionals and the use of a gradient-based approach limits the exploration and screen modeling. To overcome this limitation, the standard approach performs a survey of the different hyperparameters of the optimization problem.}
    \item \textbf{The inclusion of an additional hyperparameter increases the dimensionality of the problem, and this might reduce the efficiency of the survey strategy. This is translated into a much slower algorithm. In addition, the SO algorithm is slower than regular imaging because the screen is discretized and the power spectrum is computed, which complicates and increases the cost of the imaging problem}. As a result, the associated optimization process converges more slowly or requires more time to complete the same number of iterations as in a standard imaging problem. This challenge is particularly pronounced for large datasets. 
    \item The screen effects are complex. To understand these effects, a full marginalization would be better.
\end{itemize}

In contrast, the flexibility of multiobjective optimization enables the immediate integration of $R_{SO}$ within complex frameworks. This allows the method to leverage several advantages. These include avoiding costly hyperparameter explorations and achieving a balanced optimization with other regularizers beyond $R_{entr}$. For example, MOEA/D facilitates exploration of the Pareto front and identification of different local minima, which is particularly crucial for very sparse observations that are weakly affected by refractive noise, such as those from the EHT or ngEHT~\citep{Raymond2021, Chavez2024} at 230\,GHz. Similarly, MO-PSO offers a straightforward way to determine the marginal contribution of the regularizer, such as the screen effect on the data. In this way, the three limitations of the method proposed by~\cite{Johnson2016} were addressed.

\subsection{Synthetic data generated at 230\,GHz}\label{sec:VL}

To assess the performance of our modeling, we tested it on various synthetic structures created with \texttt{\textbf{eht-imaging}} library \citep{Chael2018}, with the assumption of static and dynamic screens at 230\,GHz. The simulated datasets include different levels of realistic gain corruptions. In the following sections, we show the results.

\textbf{There are no updates to the modeling of the screen itself in this paper, but we update the way in which the optimization problem associated with its reconstruction is solved. Our main improvement stems from the correct modeling of the multimodality of the problem, that is, of the degeneracies associated with the fact that different realizations of the scattering screen and the on-sky emission might produce similar visibilities. We identified three scenarios for which this strength can be exceptionally beneficial. First, the scattering effect at 230 GHz observations is weak even with a static source and a static screen, and it is expected to be degenerate with the noise realization. Second, when a speed is associated with the scattering, the dynamics of the screen might be misidentified with small-scale corruptions in the actual image. Finally, at 86 GHz, the screen obscures the image and significantly worsens the instrumental resolution. In this situation, small-scale structures at $20\,\mu\mathrm{as}$ scales that are typical for the EHT might be attributed to the effect of the scattering screen or the internal features of the emission. The test cases we discuss here address these three scenarios.}

\begin{figure*}
    \centering
    \includegraphics[width=0.45\linewidth]{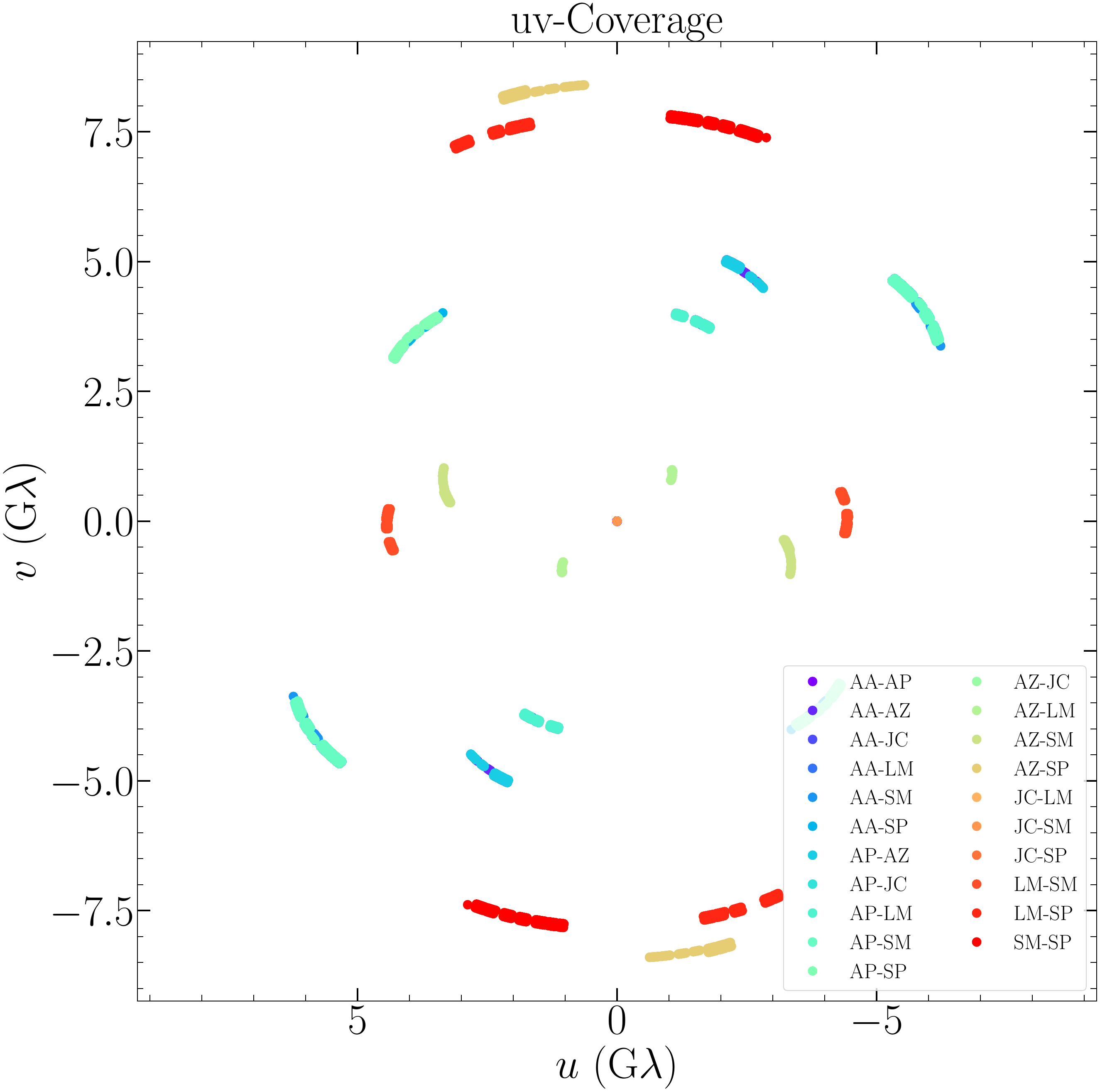}
    \includegraphics[width=0.35\linewidth]{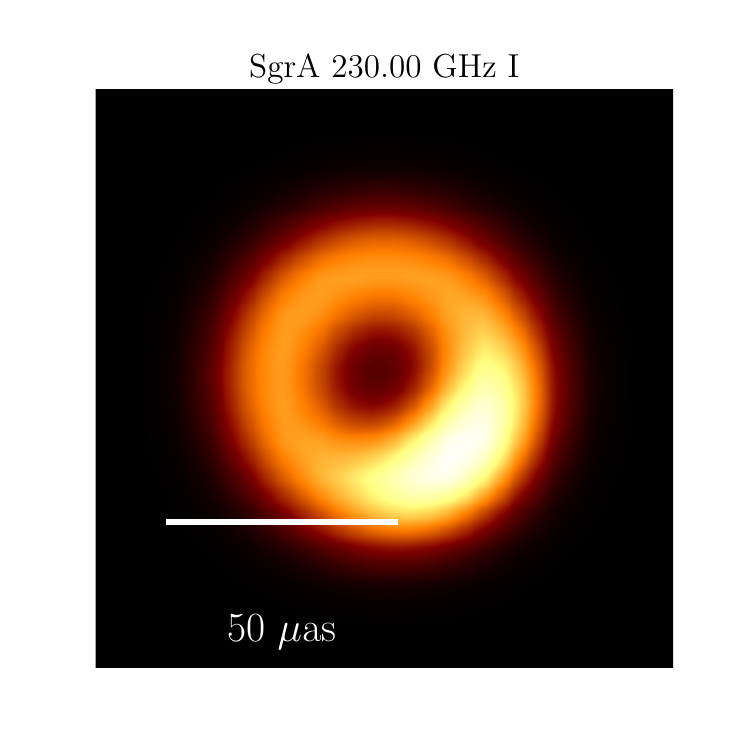}
    \includegraphics[width=\linewidth]{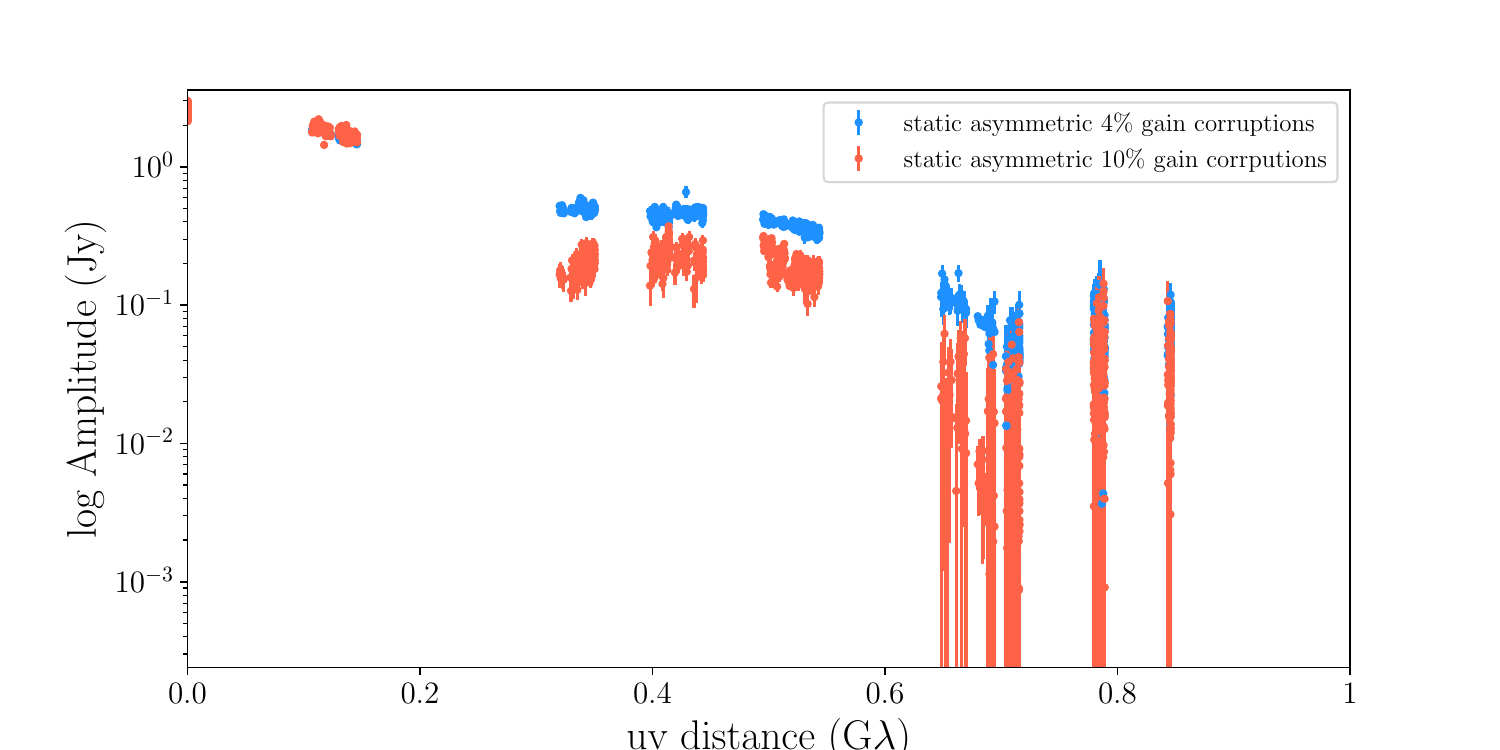}
    \caption{Top left panel: uv coverage for \sgra in the best time window of April 7, 2017. The different colors represent different baselines (indicated in the legend). Top right panel: Simulated ring in this coverage, with a brightness asymmetry. Bottom panel: Log amplitude vs. uv distance for the static scattered asymmetric ring with different gain corruptions. Blue shows 4\%, and red shows 10\%.}
    \label{fig:uv-apr11}
\end{figure*}

As a first and \textbf{simplest} example, we simulated a static ring \textbf{with an asymmetric} brightness distribution observed during the best time window (Fig.~\ref{fig:uv-apr11}, top left panel). This time window spans approximately 100 minutes and presents the best uv coverage, isotropy, and $\left(u,v\right)$ coverage fraction for \sgra in April 7, 2017, during the EHT campaign~\citep[more details can be found in][]{Farah2022, eht2022c}\footnote{The election of this window will be justified in latter sections.}. 

Gain corruptions of 4\% and 10\% were included in the data. We additionally corrupted the observations with several synthetic screens that were simulated using the \texttt{StochasticOptics}~\citep{Johnson2016} module of the \texttt{\textbf{eht-imaging}} library. The bottom panel of Fig.~\ref{fig:uv-apr11} bottom \textbf{ shows the effects of the refractive noise on the longer baselines} for the two levels of gain corruption. We demonstrate that we are able to recover the intrinsic descattered source and \textbf{the scattering screen}, and also that \textbf{the screen} prior is important in this \textbf{poorly constrained} problem.

In a second step, we increased the difficulty by simulating a population of different geometric sources that are affected by scattering screens with different velocities. We demonstrate its capability to recover the (dynamic) scattering screen even assuming a gain corruption of 10\%. \textbf{We would like to note that this scenario is only a first step toward constructing an algorithm that can adapt to real data. In EHT data, \sgra shows significant variability that stems from the evolution of the source itself and not from the dynamics of the scattering screen \citep[e.g.][]{eht2022a, Wielgus2022}. Any algorithm would need to separate the effect of the source dynamics from the effect of the scattering and the internal screen variability. Correct modeling of the dynamics of \sgra is a challenging but independent problem (even without scattering), and a wide range of solutions has been proposed \citep{Bouman2017, eht2022c, Mueller2023b, ngehtchallenge, Mus2024a, Mus2024b, Mus2024c}. We instead studied the simpler question whether we can correctly recover the speed of the screen for a static source that is scattered by a dynamically evolving screen.  Despite the simplicity of the question, this yielded two important outcomes. First, the successful demonstration of this capability helped us to constrain the currently broad range of expected screen velocities~\citep[$\sim$50 km/s to $\sim$200 km/s][]{Gwinn1991, Reid2019}. Second, it needs to be demonstrated before a framework is established that solves the more complex scenario in which the source and screen are both dynamic.}

\subsection{Recovering the screen: Algorithm}

The strategy we used is summarized in the Algorithm~\ref{alg:psopso}. First, we defined a prior image \textbf{and} a prior screen. \textbf{Throughout this paper, we understand the concept of a prior as the initial point of a minimization algorithm}. \textbf{For the case of 230\,GHz data, we used a $60\,\mu$as\footnote{From now on, we abbreviate microarcseconds by as} symmetric Gaussian as a prior. Then, we computed the wavelet coefficients, $\Psi w_I$, as described by~\citep{Mueller2022}, to help us overcome the sparsity of the data}. These inputs were used to define a composite vector $x$ that encapsulated the image and scattering information, with an overall dimension of $2\text{npix}^2 - 1$. In the first optimization step, the algorithm applied MO-PSO, using the prior image as the starting point. This optimization was carried out over $P_0$ particles and $K_0$ iterations, and it yielded an intermediate solution $x_{\text{img}}$ that was selected based on optimal regularization criteria.

The next stage involved redefining the prior as a combination of the optimized image $x_{\text{img}}$ and the initial prior scattering screen. This updated prior served as the initialization for a second MO-PSO run, which aimed to refine the solution further by simultaneously optimizing the image and the scattering screen. This stage used $P_1$ particles and $K_1$ iterations, and it ultimately produced a final image with minimized scattering effects.

\begin{algorithm}
\caption{Scattering mitigation using PSO}\label{alg:psopso}
    \begin{algorithmic}
        \Require observation uv data.
        \Require prior image (dimensions $\mathrm{npix}^2$), \textbf{wavelet coefficients}, prior screen (dimensions $\mathrm{npix}^2-1$).\\

        \State Define an image as a vector $x$ of dimension $2\mathrm{npix}^2-1$.\\

        \State Run MO-PSO using $x_0 \gets\text{prior image}$, $P_0$ particles, and $K_0$ iterations to solve the imaging problem. Solution $x_{\mathrm{img}}$ has the optimal regularizer values.\\

        \State Define $\mathrm{prior} = [x_{\mathrm{img}}, \text{prior screen}]$.\\

        \State Run MO-PSO using $x_0 \gets \mathrm{prior}$, $P_1$ particles, and $K_1$ iterations to solve the imaging problem.
        
    \end{algorithmic}
\end{algorithm}

\subsection{Results}
Throughout this paper, we evaluate the results by using the normal cross-correlation\footnote{A measure to compare two images. The closer the value to one, the more similar the two images}, or \texttt{nxcorr} metric~\cite[see][for formal definition]{eht2019d,Farah2022}. In Table~\ref{tab:nxcorr} of the Appendix~\ref{app:nxcorr}, all the \texttt{nxcorr} values for all images can be found.

\subsubsection{Different gain corruptions}

In these paragraphs, we assumed a screen of 50~km/s in th east direction during the whole observation.

\begin{figure*}
    \centering
    \includegraphics[width=\linewidth]{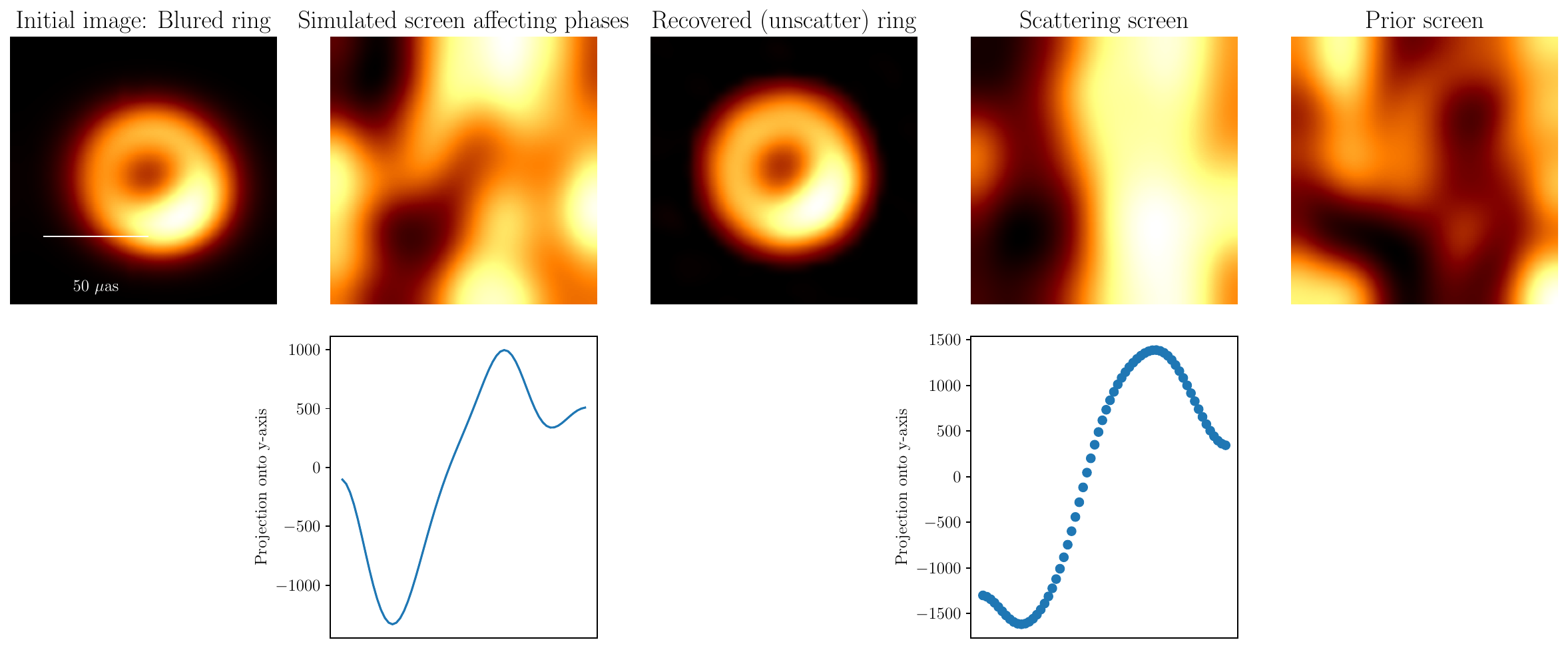}
    \includegraphics[width=\linewidth]{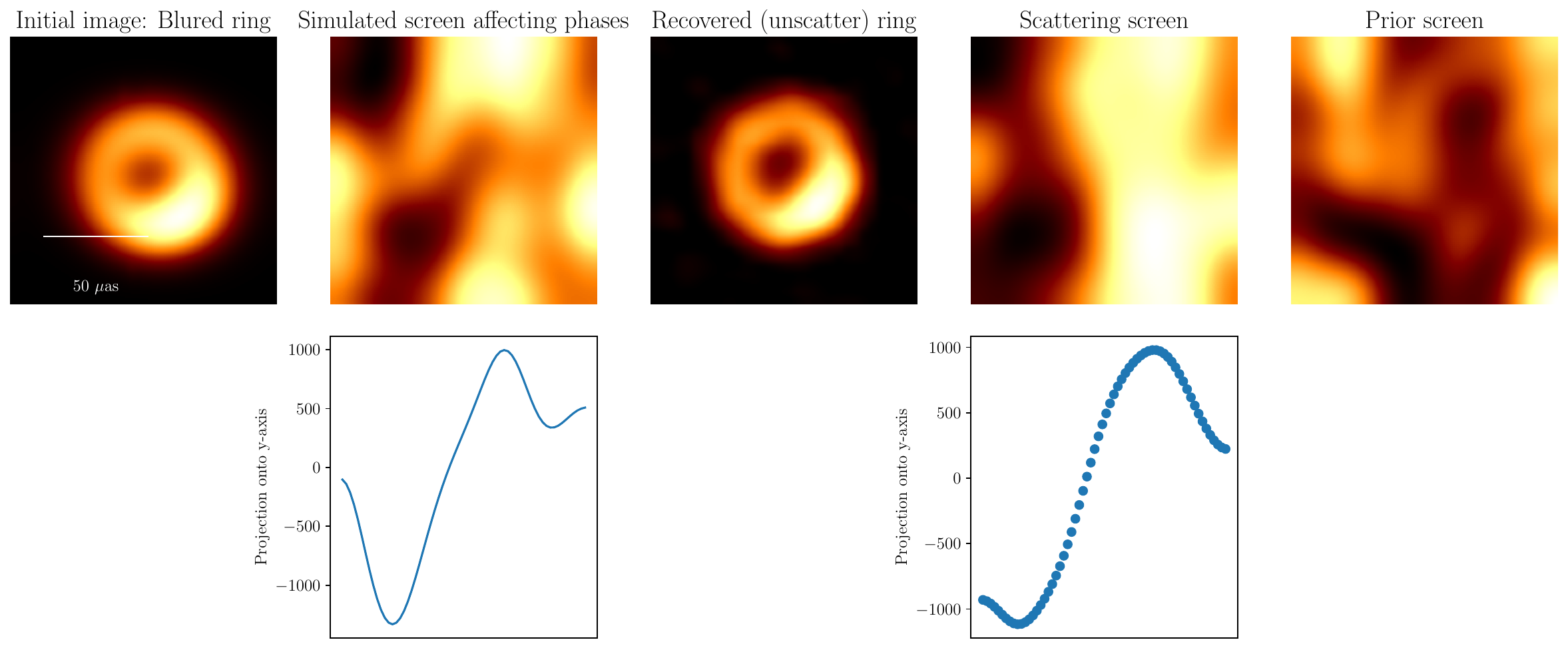}
    \caption{Scattering reconstruction using MO-PSO of the static ring with an asymmetric brightness distribution with thermal noise and a gain corruption of 4\% (top row), and with thermal noise and a gain corruption of 10\% (bottom row). From left to right: Simulated blurred ring, ground-truth screen, recovered intrinsic structure, recovered screen, and prior screen. The screens were convolved with the instrumental beam. \textbf{The subpanels below the second and fourth panels depict the projection on the y-axis of the brightness distribution along the right ascension.}}
    \label{fig:cw_scatt_4}
\end{figure*}

Figure~\ref{fig:cw_scatt_4} presents the results of the Algorithm~\ref{alg:psopso} for the cases of 4\% and 10\% gain corruptions.  Since the screen at this high frequency is poorly constrained and the \textbf{uv-coverage}  very sparse, a uniform \textbf{screen ~\cite[in contrast with the proposal of][]{Johnson2016}} is not a valid starting point.

In both cases, the recovered scattering screen are compatible, and they \textbf{show some similarities to the real screen in the sense that the positions of the brightest spots are approximately recovered and depressions are kept as well, as is visible in the projection along the y-axis of the recovered screen with the true  screen (second and fourth row). However, we would like to note that none of the finer structure in the simulated screen are recovered correctly, and the localization of the peaks and troughs is not very accurate either. This is probably to be expected because of the challenges of an array such as the EHT and the small effect of the screen at 230 GHz. Moreover, since the effect of the screen is very small at 230 GHz, the resulting, descattered image is similar to the on-sky image, that is, the on-sky image already shows a ring structure.}

\subsubsection{Independence of the screen prior}

As discussed above, \textbf{a physically feasible screen} prior is needed to recover the screen in this case with high degrees of freedom and very sparse uv-coverage. To ensure that the algorithm is not prior biased, we present two tests in which the asymmetric ring was scattered \textbf{with the same true screen}, but with two different initial screens.

\begin{figure*}
    \centering
    \includegraphics[width=\linewidth]{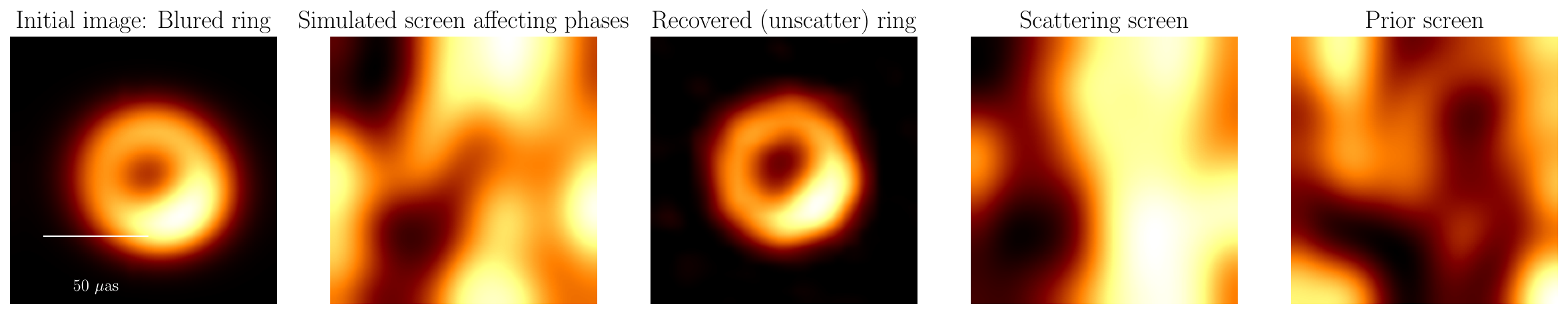}
    \includegraphics[width=\linewidth]{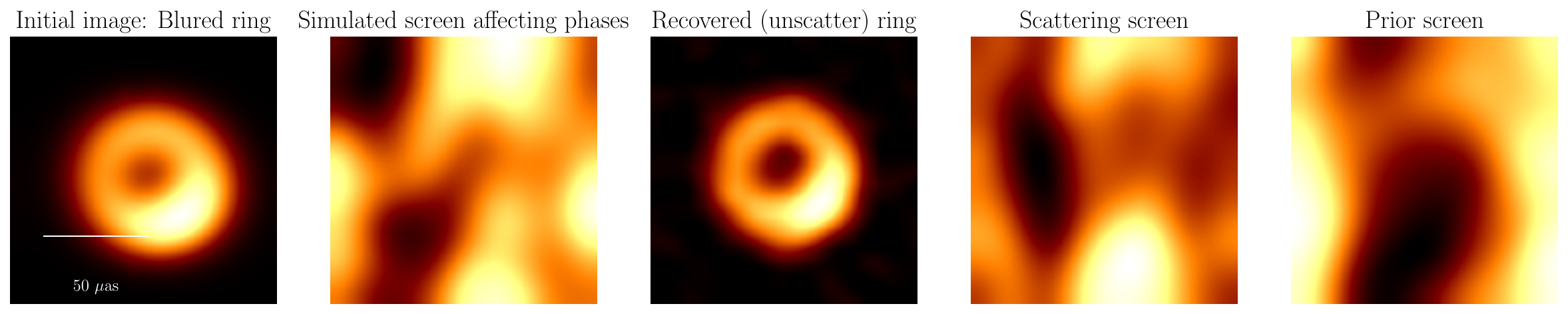}
    \caption{Comparison of the scattering reconstructions of the ring with an asymmetric brightness with different screen priors. The screens were convolved with the instrumental beam.}
    \label{fig:cw_scatt_prior}
\end{figure*}

Figure~\ref{fig:cw_scatt_prior} shows the two cases assuming systematic errors of 10\%. The reconstructed \textbf{screen is indeed clearly different}. However, the main features, such as the brightest areas and the depressions, are recovered. In particular, \textbf{the prior screen} (even if it is not similar to the real screen) can help the algorithm to converge to a better screen. Nevertheless, the refractive noise is not the dominating noise and is only weakly constrained, and the differences on the intrinsic structure reconstructions are therefore negligible. This difference is also represented in the slightly different values of the \texttt{nxcorr} shown in Table~\ref{tab:nxcorr}.

\subsubsection{Different screen velocities}
\textbf{Next, we attempted a more challenging test case. We tried to recover the speed of the scattering screen (on a static source). As discussed above, this was an artificial experiment because in practice, \sgra also evolves dynamically. However, we defer the full analysis of the source dynamics together with a moving screen to a consecutive work.} In the dynamic regime, it is essential to investigate whether the temporal evolution of the scattering screen impacts each frame of the observation. To do this, we \textbf{simulated a screen} at different velocities that affect static geometric sources based on the best time-window coverage and corrupted by thermal noise. \textbf{In particular, we simulated two screens at two different velocities, one velocity of 50\,km/s based on observations of\textbf{~\cite{Gwinn1991}}, and the other of 200\,km/s, \textbf{\cite{Reid2019}}}. 
For a scattering screen velocity of 50\,km/s, the effects of the kinematics of the screen over a period of 100\,min are minor, but a velocity of 200\,km/s produces noticeable effects. In any dynamic study, the speed of the screen therefore needs to be properly constrained. To study these effects of the speed, we solved MO-PSO + SO for every keyframe.

\begin{figure*}
    \centering
    \includegraphics[width=\linewidth]{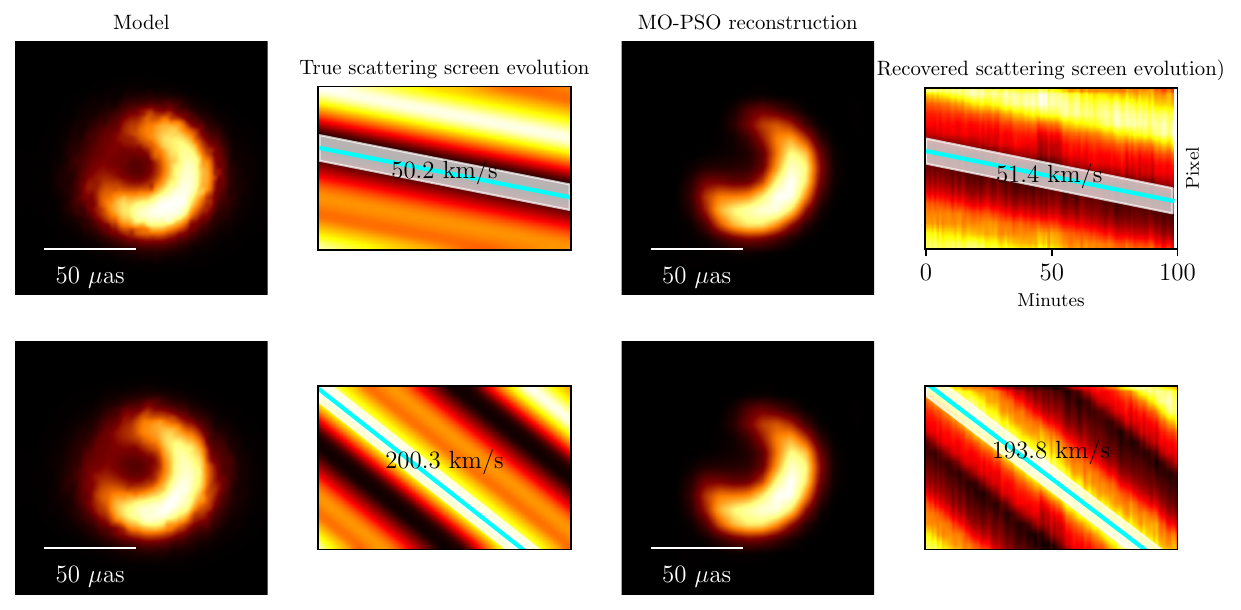}
    \caption{\textbf{Recovery of the scattering screen velocity for a screen that affects a static crescent.
    Top row: True scattering screen evolves at 50\,km/s (center left panel), while the recovered velocity is 51.4\,km/s (right panel).
    Bottom row: True scattering screen evolves at 200\,km/s (center left panel), while the recovered velocity is 193.8\,km/s (right panel). The slight deviation from the exact velocity arises from numerical errors in the fitting procedure.
    The left and center right panels show the corresponding reconstructed crescent for each case. }}
    \label{fig:screen_vel_comp}
\end{figure*}

The weak refractive noise effect on the data implies that the speed velocity does not affect the intrinsic structure recovery. Fig.~\ref{fig:screen_vel_comp} depicts a comparison of the same scattered source (first panel), a crescent, that is affected by the same screen at different velocities (upper row 50\,km/s, lower row 200\,km/s). 

To illustrate the screen evolution, we projected the pixel brightness distribution \textbf{along the y-axis} and plotted it as a function of time, following a similar approach as was described by~\citet{Mus2024a}. This is called the cylindrical plot. The third and fourth panels display the recovered intrinsic source structure and the corresponding recovered evolution of the scattering screen, respectively. \textbf{The simulated screen moved along the x-axis, so that the projection was done orthogonal to the direction of motion. The cylindrical plots (second and fourth column) can be understood as the logical extension of the projections shown in Fig. \ref{fig:cw_scatt_4}, that is, every column of the plotted matrix corresponds to the projection along the y-axis of the true and recovered scattering screen at the respective frame in time.}

\textbf{We recall from our discussion above that while the reconstruction of any structural feature finer than roughly $20\,\mu\mathrm{as}$ is not recovered, even in the static case. However, the broad overall structure of the approximate positioning of troughs and peaks across the field of view can be identified. Fig. \ref{fig:screen_vel_comp} now demonstrates that we can even track this evolution in time, and we can hence effectively measure the speed of the screen from the reconstruction by the gradient in the cylindrical plots.}

\subsubsection{Different intrinsic structures}

\begin{figure*}
    \centering
    \includegraphics[width=\linewidth]{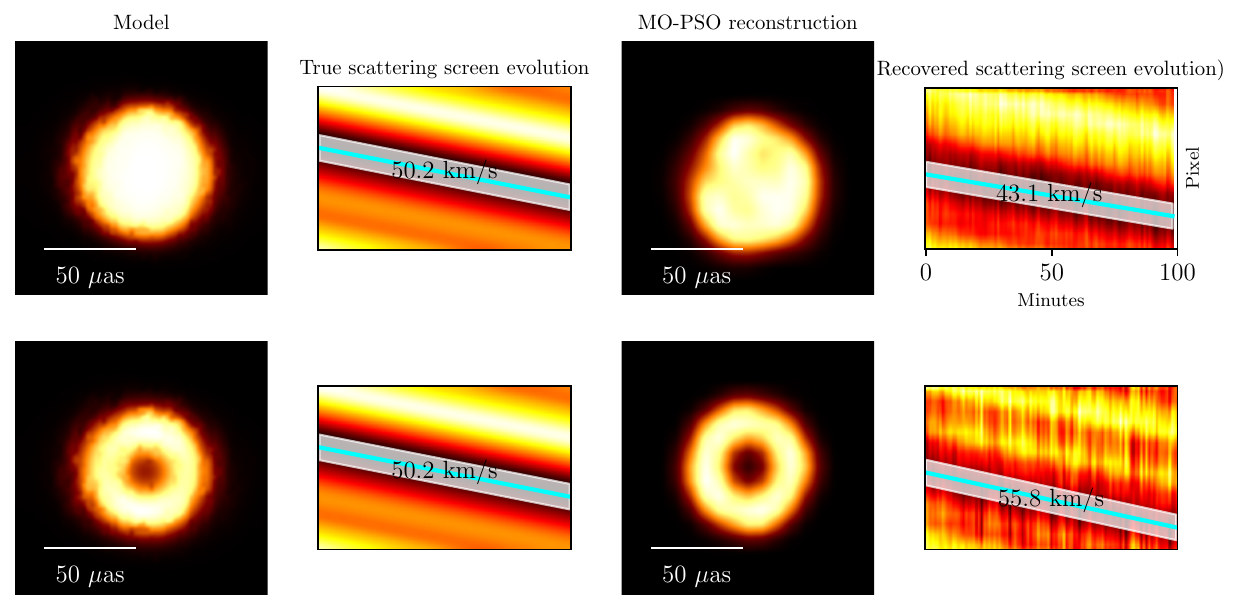}
    \caption{Dynamic scattering reconstruction of the intrinsic source and the evolving screen equivalent to that in Fig.~\ref{fig:screen_vel_comp}. Top row refers to the disk model and bottom row to the ring model.}
    \label{fig:intrinsic_struct}
\end{figure*}

Figure~\ref{fig:intrinsic_struct} presents the results of the same experiment with different observed sources. It illustrates that the geometry of the sources significantly affects the recovery of the (dynamic) screen. In the case of the disk (bottom row), the cylindrical plot for the recovered screen is noticeably less noisy than that of the ring (top row). The null in the uv-plane is affected by the screen; it is harder to solve the imaging problem. \textbf{Nevertheless, we succeeded in correctly measuring the speed of the screen, with an error smaller than $10\,\mathrm{km/s}$.}

\section{Prospects of imaging the \sgra ring at 86\,GHz}\label{sec:86ghz}

\textbf{While at 230\,GHz our method successfully recovered the overall dynamics and morphology of the screen, it was unable to capture small-scale structures~\citep[as predicted and shown in][]{Johnson2016} because the refractive noise was poorly constrained. In contrast, this noise at 86\,GHz is expected to be more constrained on longer baselines, which might allow us to recover finer structural details.}

In this section, we explore the prospects of recovering ring structures in the GMVA frequency regime. To achieve this, we exploited the multimodality capabilities of MOEA/D. We simulated observations using the 2025 GMVA configuration, incorporating Atacama Large Millimeter/submillimeter Array (ALMA) stations, including the Northern Extended Millimeter Array (NOEMA) and a double bandwidth.

\subsection{Static 86 GHz. Including gain errors}

\begin{figure*}
    \centering
    \includegraphics[width=\linewidth]{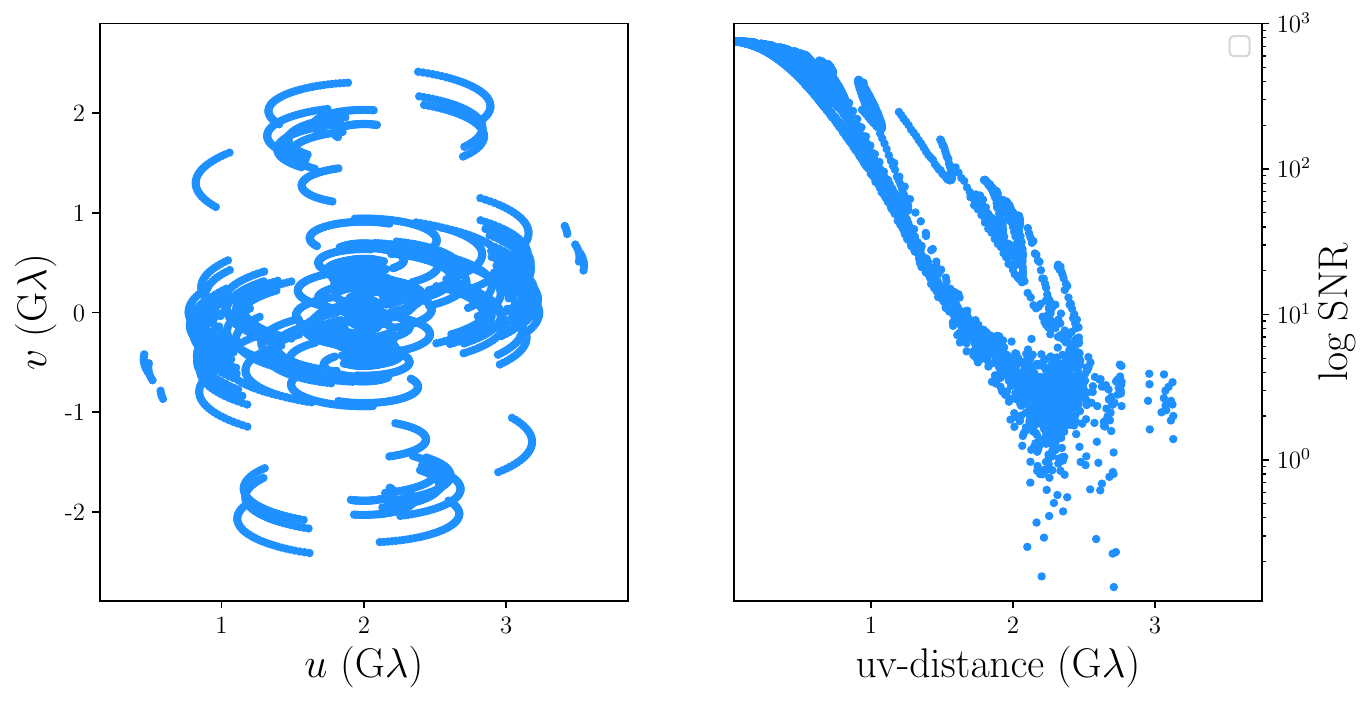}
    \caption{uv-coverage for the GMVA + ALMA array observing \sgra at 86\,GHz (left panel) and the signal-to-noise ratio in log scale vs. uv distance (right panel). The bad weather conditions and the gain corruption of 4\% cause a a low SNR even in the case of ALMA baselines.}
    \label{fig:uv_86}
\end{figure*}

\textbf{Using \texttt{ngehtsim}~\citep{ngehtsim_docs}, }we simulated static ring observations with a shadow diameter of $\sim50\,\mu$as, emulating that reported by the EHT~\citep{eht2022a}. We assumed bad weather conditions to decrease the SNR, and we included thermal errors, as well as gain errors of 4\% in phase and amplitude. The aim was to verify whether the intrinsic ring structure could be recovered. Figure~\ref{fig:uv_86} shows the uv coverage corresponding to the simulated array (left panel) and the SNR versus uv-distance (right panel).

In the range of 86\,GHz, scattering effects dominate the noise, and so the image is strongly affected. Its Strehl ratio is $S\sim 0.6$~\citep[see Fig 1 from ][]{Johnson2016}. That is to say, refractive noise at longer baselines is much more constrained. This leaves fewer degrees of freedom to reconstruct the screen. In any case, a careful mitigation strategy is crucial.

\begin{algorithm}
\caption{Scattering mitigation - Exploring multimodality}\label{alg:moeadpsoself}
    \begin{algorithmic}
        \Require observation uv data.
        \Require prior image (dimensions $\mathrm{npix}^2$), prior screen (dimensions $\mathrm{npix}^2-1$).
        \Require $n\gets$ local minima to be explored, $N\gets$ number of self-calibration iterations.
        \\

        \State Define an image as a vector $x$ of dimension $2\mathrm{npix}^2-1$.
        
        \State Create a population of $n$ images and an eight-dimensional mesh of $k$ vertices to grid the space.

        \State Define $x_0$ to be the Gaussian seen in~\cite{Issaoun2019}.\\

        \State [Optional] flag low SNR sites

        \For{$i=1,\ldots,N$}
            \State Run MOEA/D to obtain $n$ local minima.
            \State Clustering solutions by similarity: Pick the
            cluster containing more local minima, which represents the most probable solution, and call it $x_i$.
            \State Self-calibrate the visibilities with $x_i$.
            \State Recenter the image, if needed.
            \State $x_0\gets x_i$.
        \EndFor
        \\
        \State $x_{\mathrm{MOEA/D}}^{*}\gets x_N$.\\

        \State Run MO-PSO using $x_0 \gets x_{\mathrm{MOEA/D}}^{*}$, $P$ particles, and $K$ iterations. 
    \end{algorithmic}
\end{algorithm}

The complex structure of the screen and the steep brightness gradients that can occur between nearby pixels mean that genetic algorithms may struggle to converge rapidly to an optimal solution. To address this issue, we ran MO-PSO using as a starting point the cluster 
that contained the most local minima or the most probable solution, which is also the most probable solution based on the RML method \citep[although any other cluster could also be chosen; see][for further details]{Mueller2023c, Mus2024a}. By starting in a neighbor of a local minimum, we exploited gradient-based algorithms to quickly find a solution. Moreover, MO-PSO ensures convergence to the solution with the optimal regularizer contributions.

\begin{figure*}
    \centering
    \includegraphics[width=\linewidth]{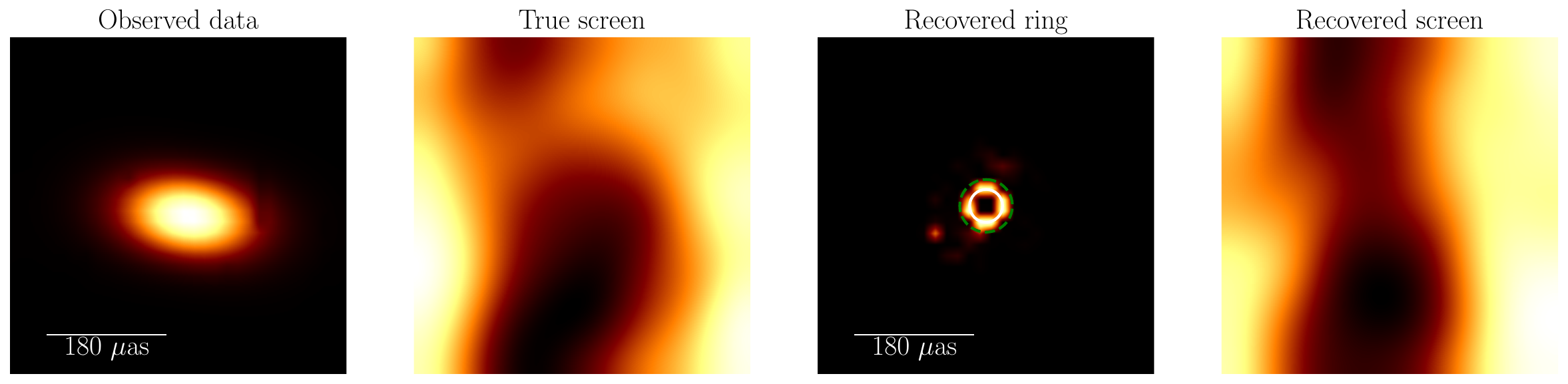}
    \caption{Scattered ring observed at 86\,GHz (first panel) with a gain corruption of 4\%. The second panel shows the simulated screen that affects the source. The third panel depicts the recovered solution using the Algorithm~\ref{alg:moeadpsoself}. The white internal circle shows the diameter of the shadow of the simulated ring (50\,$\mu$as), and the dashed green circle is 1.6 times the size of the shadow, as predicted by~\cite{Lu2023, Jong2024} in \mes. The fourth panel shows the recovered screen. The screens are blurred to the GMVA resolution. \texttt{nxcorr} can be found in Table~\ref{tab:nxcorr}.}
    \label{fig:gmva_sol_good}
\end{figure*}

The strategy we used is described below and summarized in Algorithm~\ref{alg:moeadpsoself}. We initialized the MOEA/D algorithm with a population of 340 individuals and 51 neighbors. For the entropy prior, we assumed a uniform Gaussian  of $60,\mu$as. The first generation of individuals was initialized with a Gaussian whose beam parameters were taken from the least-squares (LSQ) Gaussian fit reported by~\cite{Issaoun2019}, with values of $\left(120,\mu\mathrm{as}, 100,\mu\mathrm{as}\right)$ and a position angle (PA) of 90 degrees. We flagged \textbf{data} with an SNR below 10, which is equivalent to approximately 1.25\% of the maximum SNR. We considered the data terms of MOEA/D to be closure quantities (phases and log-amplitudes). We let the population evolve over 1000 generations. After the population evolution was complete, we clustered the 340 individuals (solutions) by similarity, requiring 10\% similar features to belong to the same cluster. We selected a representative image from the cluster that contained more solutions (the most probable local minimum),
we self-calibrated $N$ times, running MOEA/D in every iteration, and we used this image as a starting point and let the new population (again composed of 340 individuals and 51 neighbors) evolve over 500 generations.\footnote{We ran 500 generations to speed the process up.}. When it was required, we recentered the source to the (0,0) since the relative position information can be lost due to the use of closures. Fig.~\ref{fig:paretos_86_allFalse} shows the set of clusters for three iterations. The clusters surrounded by a green box have the highest \texttt{nxcorr} with respect to the model. The red box marks the cluster that contained more solutions, and the blue box shows the cluster with better hyperparameter combination.
The solution we obtained after the last self-calibration iteration was used as starting point to solve MO-PSO. The faster convergence of MO-PSO allowed us to obtain a more accurate solution faster.
Fig.~\ref{fig:gmva_sol_good} shows the final source and screen reconstruction. The intrinsic structure has a \texttt{nxcorr} of 0.97, and the nxcorr of the recovered screen is 0.81 (blurred) and 0.88 (nonblurred). These metrics demonstrate how successful the algorithm was in recovering a screen in a more constrained case.

In the next section, we apply MOEA/D with different starting points and report the probabilities of obtaining a ring solution under optimal conditions using a priori calibrated data. Our findings indicate that for certain initial conditions, no convergence to a ring solution is reached, and the probability for others, such as a Gaussian starting point, remains low, but a ring can indeed be recovered.

\subsection{Static 86 GHz. A priori calibrated data with thermal noise}\label{app:scatt_freq}

In this section, we explore the chances of obtaining a ring structure using different starting points when the a priori calibrated data are only corrupted with thermal noise. The goal of this section is to demonstrate \textbf{the importance of the global exploration and }that the assumed starting geometry may condition the problem, so that no ring can be recovered even in favorable situations. The strategy we used is summarized in Algorithm~\ref{alg:moeadpsonoself}.
The three starting points were the Gaussian observed by~\cite{Issaoun2019} using LSQ before applying SO (major FWHM of 228\,$\mu$as, minor FWHM of 143\,$\mu$as and PA of 86 grad), a disk of 60\,$\mu$as, and a ring with the expected shadow size seen by the EHT of $50\,\mu$as\textbf{~\citep{eht2022a}}. Fig.~\ref{fig:x0_86} shows the images (top row) and their amplitude versus uv distance in log scale (bottom row).

For every case, we explored $\sim360$ local minima and let the population to evolve during 10000 generations. We allowed each individual to interact (crossover) with $15\%$ of the population. The data-term functional was the $\chi^2$ of the visibilities.

\textbf{Figure~\ref{fig:paretos_86} presents the Pareto fronts of the solutions originating from three distinct starting points, the Gaussian, disk, and ring, each clustered using a uniform similarity threshold of $10\%$. The title of each cluster indicates the percentage of solutions contained within it, along with the $\chi^2$ value, which quantifies the discrepancy between the true unscattered ring and the reconstructed image.} 

The different number of clusters contained in every starting point is the first thing to note. The starting point conditions the diversity of the recovered solutions. Then, the ring starting from a Gaussian can be recovered, but not the ring from a disk. This means that the optimization algorithm tends to find disk structures faster than a ring. Therefore, the optimization needs to start from a distant point. 
Another interesting fact is $\chi^2$ is systematically lower in all the clusters with ring-like structures. Therefore, even though the remaining clusters may approximate the Pareto front (i.e., are local minima, valid images), those with a ring present the best fit to the data.

Table~\ref{tab:summ_moead} summarizes the results of the MOEA/D exploration with the different starting points. For the Gaussian starting point, 25.45\% of the solutions are ring-like ($\sim85$ local minima). This means that the probability of finding a local solution that represents a ring is just $\sim0.25$. Therefore, unimodal optimization methods (based on gradients or Hessian) are likely to find nonring solutions. In contrast, using a ring with the expected size of \sgra as the initial point increases the probability to almost to 40\%. Although considering a ring as initial point maybe a strong bias, we can claim that with two completely different starting points, MOEA/D is able to recover a ring. 

\textbf{Thus, we conclude that the choice of prior complicates the accurate recovery of the true source structure. In this context, standard local searches struggle to explore alternative candidates, and they therefore are suboptimal strategies for identifying lower-probability solutions in a highly ill-posed problem.}

\begin{table*}[h]
    \centering
    \caption{Clustering outcomes of MOEA/D initial populations by morphology}
    \begin{tabular}{l c c c}
        \toprule
        & Number of clusters & Number of ring clusters & Total ring solutions (\%) \\
        \midrule
        Gaussian &  9&  2&   25.45  \\
        Disk  &   4&  0&   0 \\
        Ring  &   5&  2&   37.57 \\
        \bottomrule
    \end{tabular}
    \tablefoot{Overview of clustering results for initial points used in the MOEA/D algorithm. The table shows the total number of clusters identified with a similarity threshold of 10\%, the number of the clusters that present a ring structure, and the percentage of the total solutions (out of 340) that exhibit ring-like morpohologies.}
    \label{tab:summ_moead}
\end{table*}

\begin{algorithm}
\caption{Scattering mitigation - Exploring multimodality without self-calibration}\label{alg:moeadpsonoself}
    \begin{algorithmic}
        \Require observation uv data.
        \Require prior image (dimensions $\mathrm{npix}^2$), prior screen (dimensions $\mathrm{npix}^2-1$).\\

        \State Define an image as a vector $x$ of dimension $2\mathrm{npix}^2-1$.
        
        \State Create a population of $n$ images and an eight-dimensional mesh of $k$ vertices to grid the space.\\

        \State Run MOEA/D to obtain $n$ local minima.\\
        
        \State Clustering solutions by similarity: Select the cluster with fewer $\chi^2$, $x_{\mathrm{MOEA/D}}^{*}$.\\

        \State Run MO-PSO using $x_0 \gets x_{\mathrm{MOEA/D}}^{*}$, $P$ particles and $K$ iterations. 
    \end{algorithmic}
\end{algorithm}

\subsection{Conclusions}

\textbf{We have investigated the feasibility of recovering the \sgra ring structure at 86\,GHz in the presence of refractive scattering and various observational challenges. The refractive noise at this lower frequency, although dominant, is more strongly constrained on longer baselines than at 230\,GHz. This allowed our global optimization schemes (MOEA/D followed by MO-PSO) to resolve finer structural features. Even under adverse conditions, such as a low signal-to-noise ratio and gain errors, the proposed multistep strategy robustly recovers the intrinsic source geometry and a meaningful approximation of the scattering screen. The final \texttt{nxcorr} values reach 0.97 for the source.}

\textbf{Moreover, we demonstrated that the choice of the initial image geometry significantly influences whether the optimization converges to a ring solution. In idealized simulations with a priori calibrated data, starting from a Gaussian prior yields a probability of approximately 25\% of recovering a ring, whereas beginning with a ring prior increases this probability to nearly 40\%. By contrast, a disk prior never converged to a ring solution in our tests. These findings underscore the importance of genuinely global exploration methods and of the use of diverse initial conditions when strongly scattered sources such as \sgra are imaged. Overall, a comprehensive multimodal approach is essential to mitigate the ill-posed nature of VLBI data at 86\,GHz and to reliably recover ring-like structures in the presence of refractive scattering.}

\section{Marginal contribution of the SO regularizer}\label{sec:marginal}

To conclude the analysis, we examined the marginal contribution of the Regularizer~\eqref{eq:SO} at frequencies of 230 GHz and 86 GHz.
We followed a strategy similar to that outlined by~\cite{Mus2024c}. We ran MO-PSO with the same initial point, number of iterations, simulated screen, and number of particles, but varied the initialization of these particles by following a uniform distribution. This approach allowed us to assess the impact of the regularizer for different refractive noise conditions. Using MO-PSO, we can determine the relative importance of a given regularizer on the final reconstruction. This information constrains the possible values of the regularizer. The lower the variance in the values, the more constrained and significant the regularizer for the reconstruction.

\begin{figure}
    \centering
    \includegraphics[width=\linewidth]{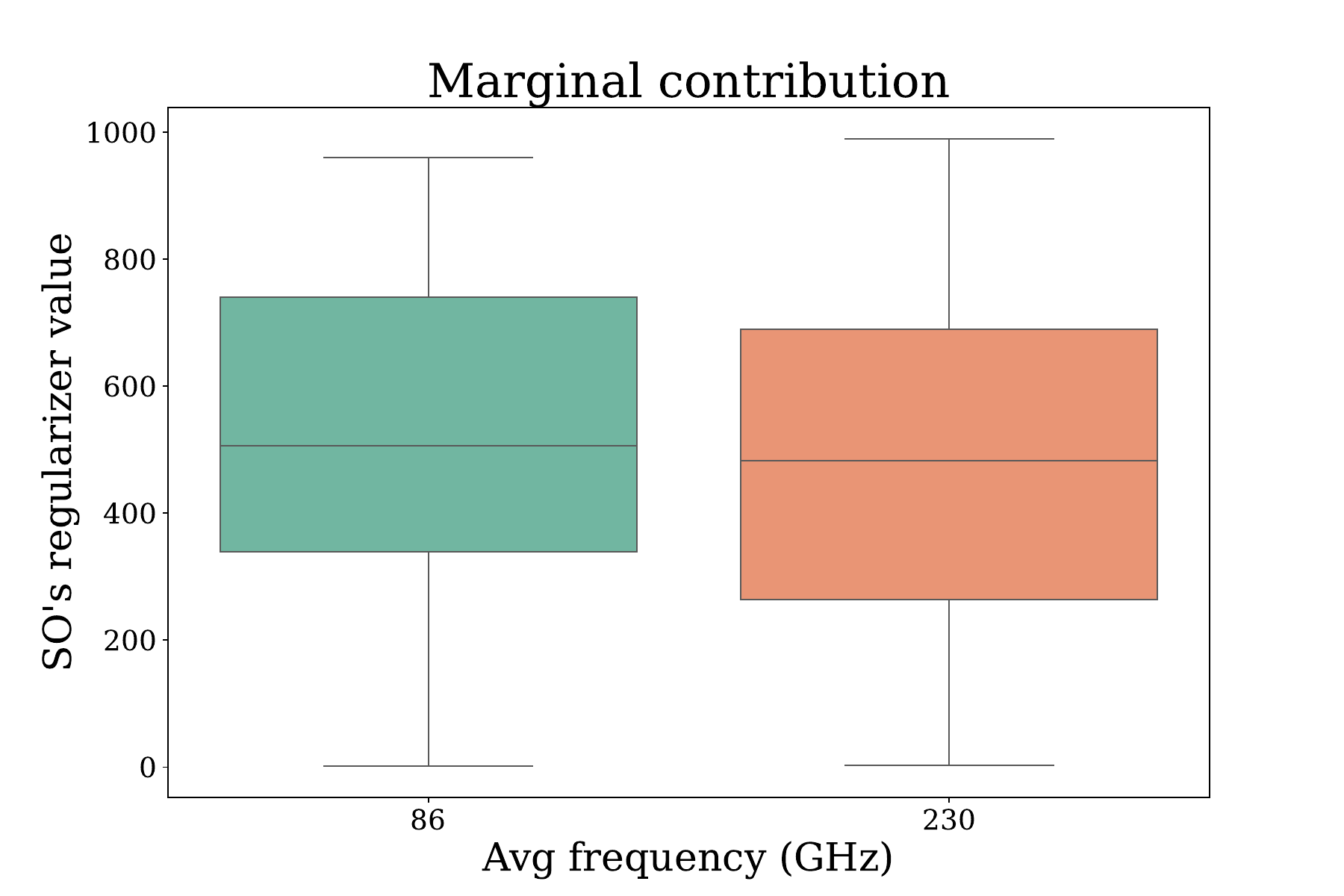}
    \caption{Marginal contribution of the hyperparameter associated with Eq.~\eqref{eq:SO} for observations at 86 and 230\,GHz.}
    \label{fig:marginal_contribution}
\end{figure}

\begin{table*}
    \caption{Summary statistics of the marginal-contribution regularizer}
    \centering  
    \begin{tabular}{lcccc}
    \toprule
     & mean & std & min & max \\
    Frequency (GHz) &  &  &  &  \\
    \midrule
    86 & 521.175708 & 248.024702 & 1.600457 & 960.440496 \\
    230 & 490.462748 & 282.973316 & 3.254598 & 989.571094 \\
    \bottomrule
    \end{tabular}
    \label{tab:marginal_contribution}
    \tablefoot{Mean, standard deviation, minimum, and maximum values for the SO-regularizer hyperparameter in Eq.~\eqref{eq:SO}. computed over all trials. At 86 GHz, the hyperparameter is more tightly clustered (lower variance) than at 230 GHz, indicating a more constrained regularization regime.}
\end{table*}

Figure~\ref{fig:marginal_contribution} shows the distribution of the hyperparameter values for both frequencies. As expected, we note a broader distribution for 230\,GHz, meaning that the regularizer is less constrained. Table~\ref{tab:marginal_contribution} summarizes the main features. The standard deviation (std) for 230\,GHz is indeed larger than for 86\,GHz

\section{Future work}\label{sec:future}

This work opens several promising directions for future investigation. First, the \texttt{nxcorr} metric was found to be suboptimal for validating our results, as elaborated in Appendix~\ref{app:nxcorr}. While the main features of the reconstructions may be recovered, the metric values often do not accurately reflect the visual quality of the outcomes. In future works, we will aim to find a better and appropriate metric that successfully reflects the \textbf{fidelity} of the reconstructions.

Second, we considered \sgra as a static source. However, the reconstruction of the scattering screen in \sgra may present an \textbf{additional} challenge due to the intrinsic variability on minute timescales of this source~\citep{eht2022c}. A wide range of movie-making algorithms that account for time-domain correlations have been proposed \citep{Bouman2017, Mueller2022, ngehtchallenge, Mus2024a, Mus2024b, Mus2024c}. \textbf{With the recent studies of the dynamics of \sgra at horizon scales} (EHTC, in prep.), future research will focus on incorporating \sgra as a dynamic source and a dynamic scattering screen to achieve reconstructions of real data by better capturing the time-dependent behavior of the system. 

Third, we currently investigate how the intrinsic geometry of the initial conditions influences the multimodality exploration. We observed that starting the exploration with certain geometric models, for example, disk, prevented us from recovering the intrinsic ring, while using a Gaussian as an initial point allowed us to recover ring structures for $\sim25$\% of the cases (see Sect~\ref{app:scatt_freq} for further details). Another case we aim to understand is why the reconstruction of the scattering screen appears to be noisier when applied to a ring structure, as indicated in Fig.~\ref{fig:screen_vel_comp}.

Fourth, we intend to apply this strategy to real data at 86\,GHz that are observed with the new sites and capabilities of the GMVA+ALMA array, and we will try to obtain the real ring of \sgra.

Last, we currently examine the effects of scattering on hotspot tracking. This might provide deeper insights into the complex dynamics of the system.

\section{Summary and conclusions}

\textbf{Scattering remains one of the most significant challenges in radio astronomy observations of the Galactic Plane, particularly at low frequencies. It complicates the image reconstruction by introducing substantial stochastic distortions that can severely degrade the quality of the recovered images~\citep{JohnsonNarayan2016} and corrupt signals. This poses considerable difficulties for tasks such as pulsar searches~\citep{Narayan1992}.}

Several methods have been developed to mitigate its effects~\citep{Fish2014,Johnson2018} and even to reconstruct the affecting screen. However, these methods are constrained by their tendency to find only local minima and are limited to a restricted set of regularizers due to constraints of the computational power. 

We introduced a new strategy to mitigate scattering and reconstruct the screen that integrates the modeling of stochastic optics within the framework of a multiobjective optimization. \textbf{Our method provides a mathematically rigorous solution to the problem without being restricted to a specific screen model or intrinsic source structure. This flexibility is particularly valuable to model the highly nonlinear and ill-posed nature of the VLBI imaging and screen. In contrast to traditional noise-inflation and deblurring techniques~\citep{eht2022c, eht2024b} and to standard unimodal exploration~\citep{Johnson2016}, our strategy thoroughly explores the family of local solutions. This crucial aspect is frequently overlooked in conventional approaches.}
\textbf{In this way, }this approach solves the challenges faced by previous methods, such as 1) the high computational cost, 2) the inefficiency due to the increased dimensionality of the problem by the inclusion of a new regularizer, and 3) the loss of information about the effects of scattering mitigation on the image. 
By solving Prob.~\eqref{prob:mop_ours_scalar} using nature-inspired optimization techniques (specifically, genetic algorithms and particle swarm optimization), we effectively explored the multimodal nature of the problem.

We demonstrated that this novel strategy successfully mitigates scattering in very sparse EHT observations of approximately 100\,minutes at 230\,GHz, where scattering noise is weak and poorly constrained, and in high SNR GMVA+ALMA observations at 86\,GHz that lasted about 12\,hours, where scattering is stronger and better constrained.

We applied our algorithm to \textbf{synthetic} EHT observations to test various screen velocities, intrinsic geometric sources, levels of gain corruption, and different priors. In all cases, the recovered unscattered source and the corrupting scattering screens were well reconstructed. \textbf{Notably, even under these poorly constrained conditions, our method succeeded in recovering the screen velocity. This is an improvement over earlier approaches, and it is crucial for constraining the expected range~\citep{Gwinn1991, Reid2019}}. However, in this high-frequency domain, the prior screen proved crucial for aiding convergence, as expected because the nature of the problem is poorly constrained. 

Additionally, we modeled a ring structure and applied the scattering kernel predicted by~\citet{Bower2015, Psaltis2018} for \sgra at 86\,GHz. This introduced various gain corruptions, and we consistently assumed adverse weather conditions. 
Our objective was to recover the intrinsic ring structure despite these challenges. Leveraging the significant exploration capabilities of genetic algorithms, we were able to successfully reconstruct the intrinsic ring and the scattering screen.

To this end, we proposed two different strategies based on the level of gain corruption under two representative gain corruption conditions. One strategy was designed to only explore multimodality, and the second strategy selected the most probable solution and self-calibrated to it. The strategy iterated this procedure as many times as imposed by the user.

We analyzed the different local minima we recovered and determined, many of which them had a ring morphology based on different initial configurations (Gaussian, ring, and disk). Our findings indicate that, for instance, starting from a disk configuration presents the highest likelihood of failure to accurately recover the ring. 
Because of the more constrained nature of this problem, we were able to recover a better image of the scattering screen.

We finally explored the marginal contribution of the stochastic optics regularizer. \textbf{Our findings confirmed that at lower frequencies, where long baselines are dominated by refractive noise, the constraints are tighter. This reduces the degrees of freedom of the regularizser.}

In conclusion, multiobjective optimization together with exploration algorithms is a tool that might enhance our ability to see the ring of \sgra in real GMVA+ALMA observations at 86\,GHz. It has not been detected at this frequency so far.

\begin{acknowledgement}
This work was supported by the Italian Ministry of University and Research (MUR)– Project CUP F53D23001260001, funded by the European Union – NextGenerationEU. T.T acknowledge the ``Center of Excellence Severo Ochoa" grant CEX2021-001131-S funded by MCIN/AEI/ 10.13039/501100011033 awarded to the Instituto de Astrof\'{\i}sica de Andaluc\'{\i}a. H.M, G.Y and A.L. acknowledge the M2FINDERS project funded by the European Research Council (ERC) under the European Union’s Horizon 2020 Research and Innovation Programme (Grant Agreement No. 101018682) and by the MICINN Research Project PID2019-108995GB-C22. The National Radio Astronomy Observatory is a facility of the National Science Foundation operated under cooperative agreement by Associated Universities, Inc.
The authors gratefully acknowledge Y. Kovalev for his insightful comments and valuable suggestions, and the anonymous referee for their constructive feedback.
\\
\\
\textbf{Data Availability}\label{sec:softwareAvailability}
Our imaging pipeline and our software is included in the second release of MrBeam \footnote{\url{https://github.com/hmuellergoe/mrbeam}}. Our software makes use of the publicly available \textbf{eht-imaging} library \citep{Chael2016, Chael2018}, regpy \citep{regpy}, MrBeam \citep{Mueller2022, Mueller2023a, Mueller2023b, Mueller2023c} and pyswarms \footnote{\url{https://pyswarms.readthedocs.io/en/latest/}} packages.
\end{acknowledgement}

\bibliographystyle{aa}
\bibliography{lib}{}

\begin{appendix}
\onecolumn
\section{Additional figures}

\begin{figure}[h!]
    \centering
    \includegraphics[width=\linewidth]{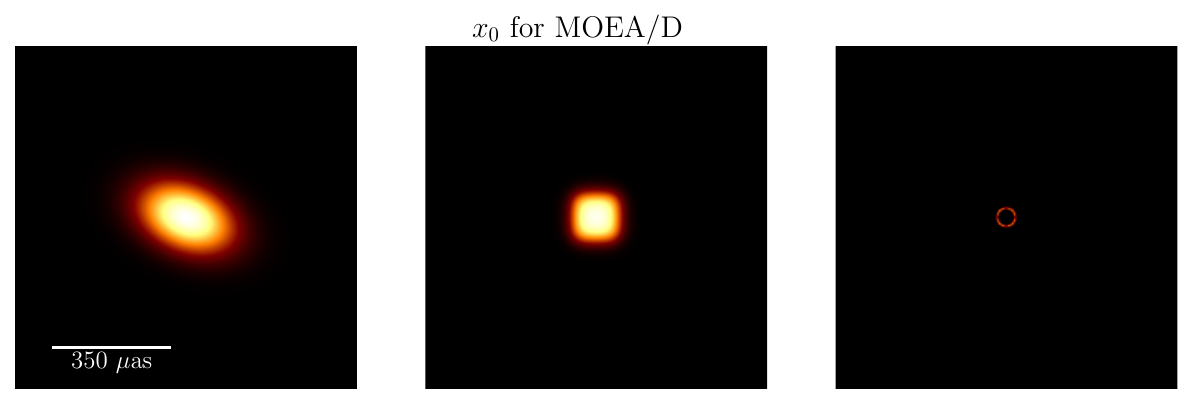}
    \includegraphics[width=\linewidth]{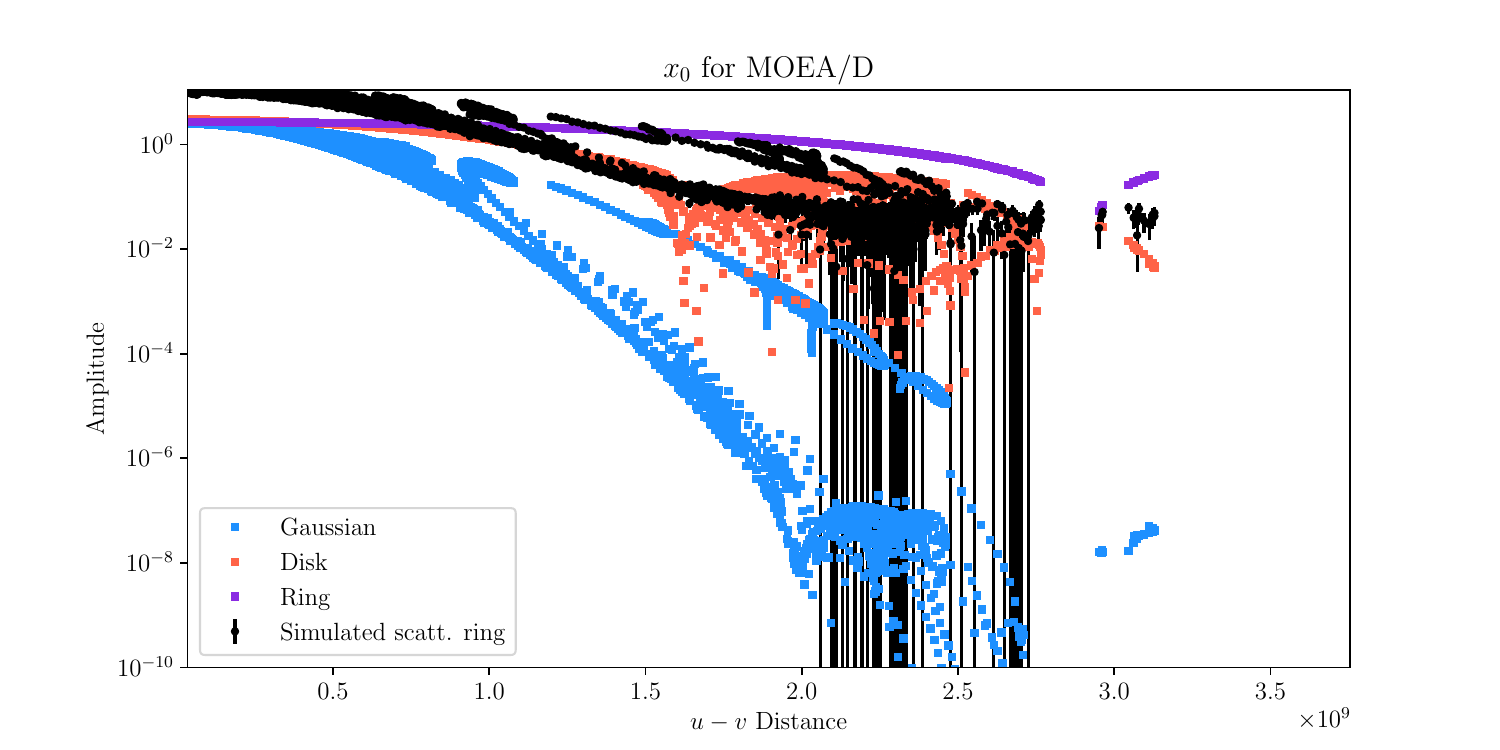}
    \caption{To explore the multimodality of the solutions at 86\,GHz we set the starting point for MOEA/D to be a Gaussian~\citep[corresponding to LSQ solution in][]{Issaoun2019}, disk and ring. Top row of the figure shows the images, bottom row, the amplitude vs uv-distance including the simulated scattered ring (black dots).}
    \label{fig:x0_86}
\end{figure}

\section{Comparison metrics}\label{app:nxcorr}

In this section, we present the \texttt{nxcorr} values obtained from all the reconstructions performed. Table~\ref{tab:nxcorr} contains the \texttt{nxcorr} values of the reconstructions of the intrinsic descattered source and the recovered (blurred and non-blurred) scattering screen appearing in Figures~\ref{fig:cw_scatt_4},~\ref{fig:cw_scatt_prior} and~\ref{fig:gmva_sol_good}. It is important to note that the \texttt{nxcorr} tends to improve at lower frequencies, which is an expected outcome due to the increased constraints at those frequencies. However, it should also be noted that the \texttt{nxcorr} may not be the most suitable metric for evaluation. The \texttt{nxcorr} assumes a Pearson correlation, which may not be optimal for non-normally distributed data and because of the given the high variance in pixel brightness values (ranging between -3000 to 3000 radians in some instances), which significantly reduces the accuracy of the comparisons.
\begin{equation}\label{eq:nxcorr}
    \rho_{\text{NX}}(X, Y) = \frac{1}{N} \sum_i \frac{(X_i - \langle X \rangle)(Y_i - \langle Y \rangle)}{\sigma_X \sigma_Y}.    
\end{equation}

Eq.~\eqref{eq:nxcorr} provides the definition of \texttt{nxcorr}, used to compare two images, $X$ and $Y$. Due to the high variance in this context, the application of the mean and normalization by the standard deviation ($\sigma$) may be less appropriate. As discussed in Sect.~\ref{sec:future}, we are developing a more sophisticated metric to address these limitations.

\begin{table}[h!]
    \centering
    \caption{\texttt{nxcorr} values for the different reconstructions presented.}
    \begin{tabular}{l c c c}
        \toprule
        & \texttt{nxcorr} intrinsic source & \texttt{nxcorr} blurred screen &  \texttt{nxcorr} non-blurred screen\\
        \midrule
        Figure~\ref{fig:cw_scatt_4} &  0.99 (top row); 0.99 (bottom row) & 0.55 (top row); 0.57 (bottom row) & 0.73 (top row); 0.75 (bottom row) \\
        Figure~\ref{fig:cw_scatt_prior}  & 0.99 (top row); 0.99 (bottom row) & 0.57 (top row); 0.45 (bottom row) & 0.75 (top row); 0.55 (bottom row)\\
        Figure~\ref{fig:gmva_sol_good} &   0.97 & 0.81 & 0.88 \\
        \bottomrule
    \end{tabular}
    \label{tab:nxcorr}
\end{table}

\section{Pareto fronts}\label{app:paretofronts}

In this appendix we show the Pareto fronts of the different sources discussed in the main text. 
\begin{figure}[h]
    \centering
    \includegraphics[width=.3\linewidth]{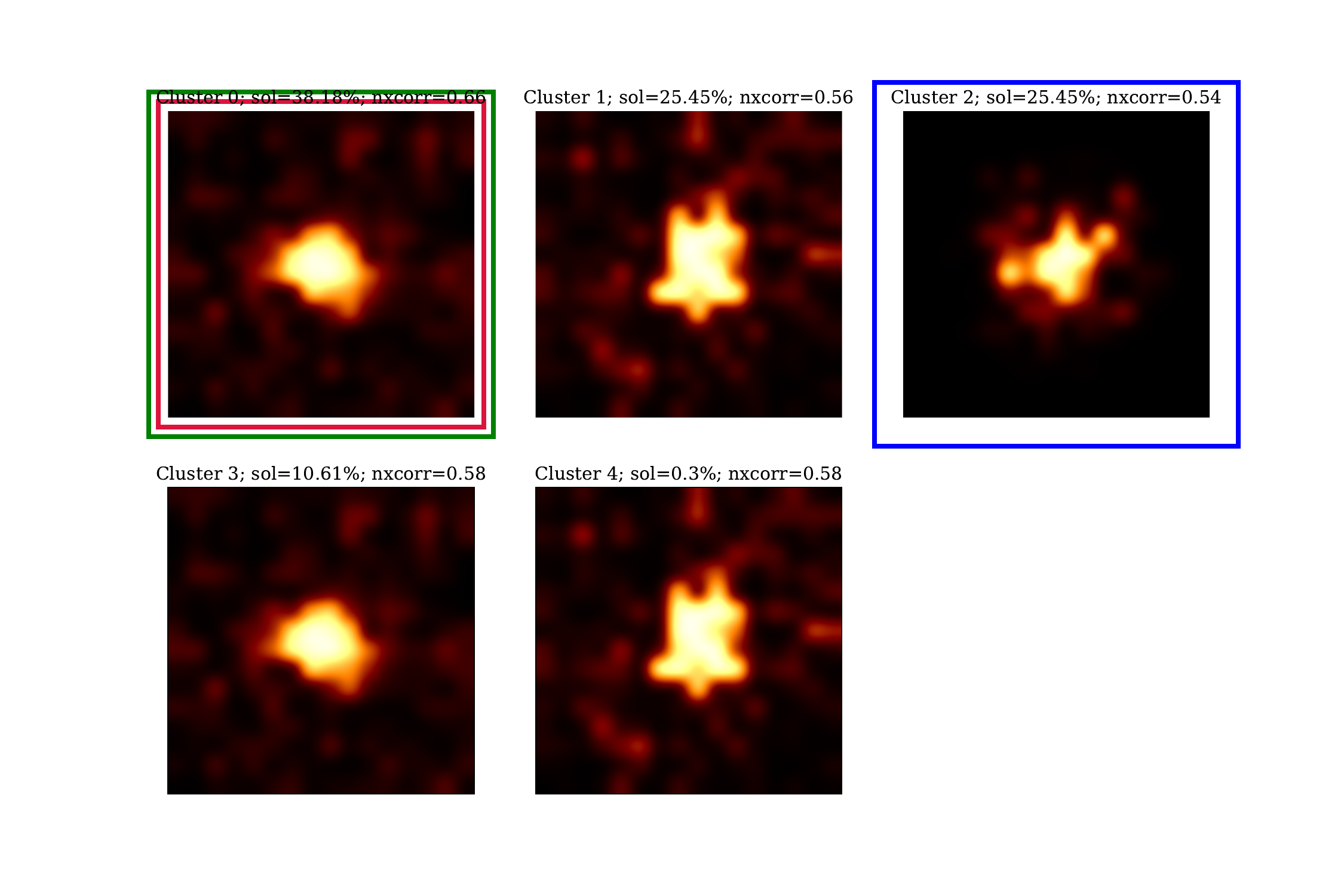}
    \includegraphics[width=.3\linewidth]{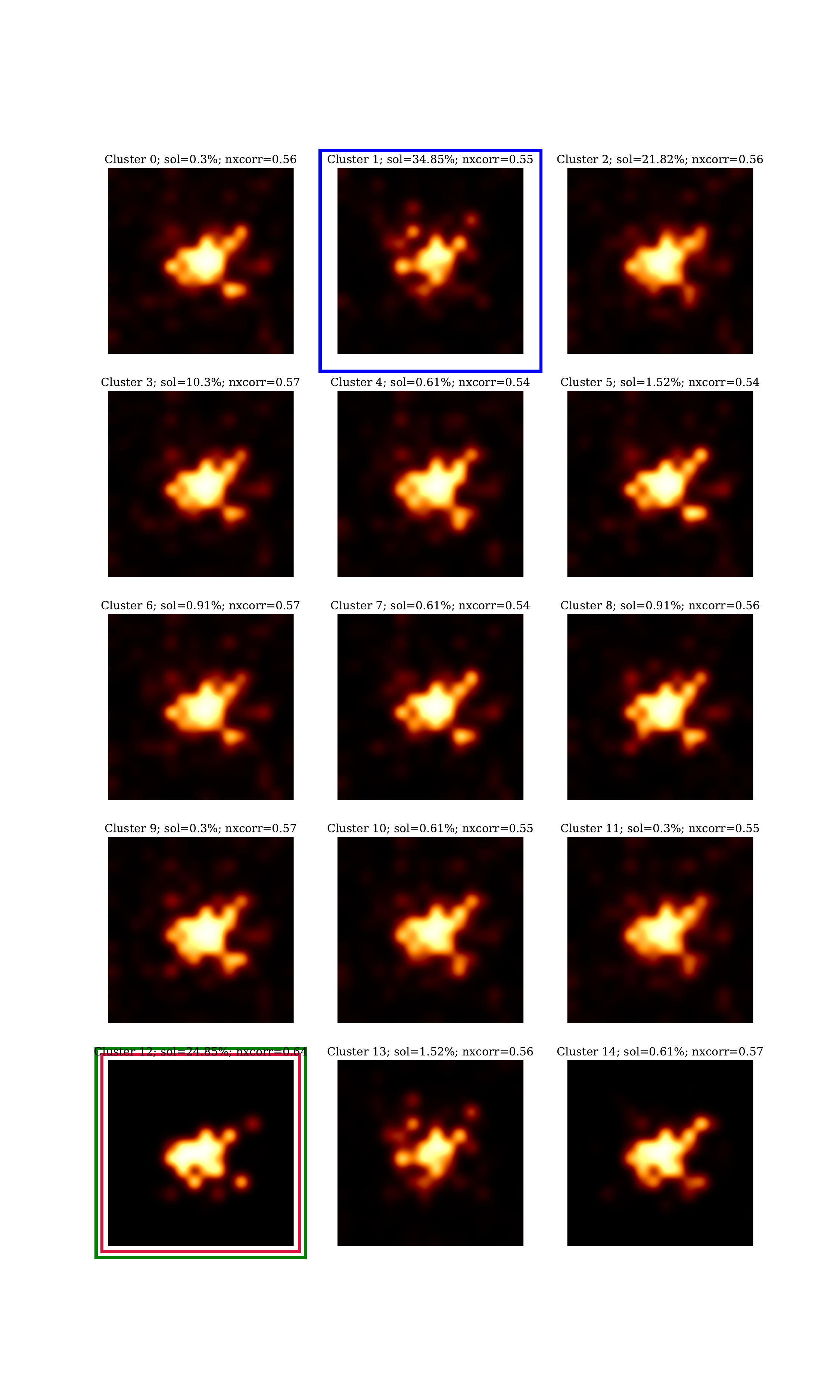}
    \includegraphics[width=.3\linewidth]{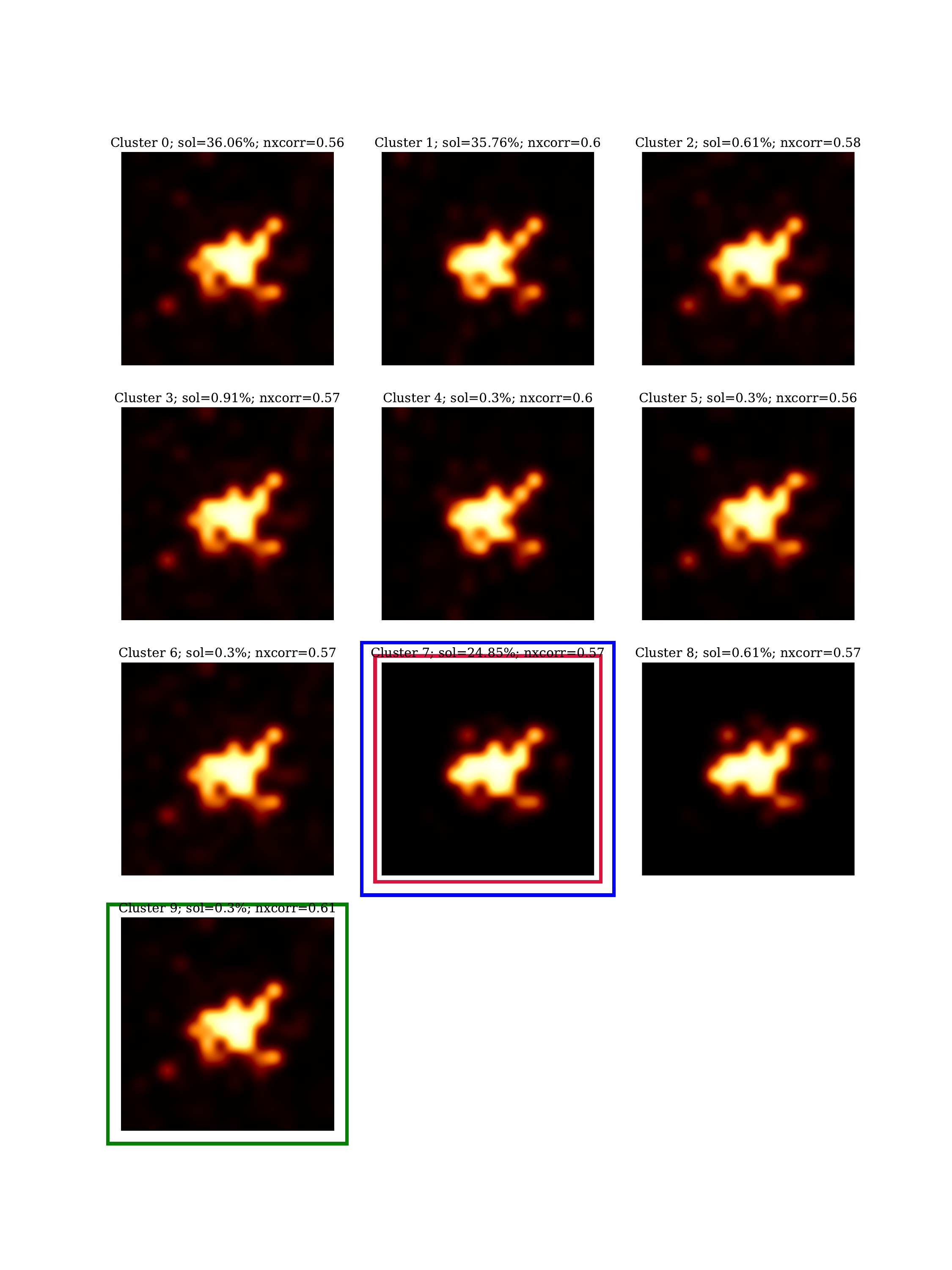}
    \caption{ Pareto front for the intrinsic source at 86\,GHz across each self-calibration iteration. Arranged from left to right, the panels correspond to the first, second, and third iterations, respectively. In the initial iteration, the ring structure is absent from the solution set. A hint of the ring appears in the second iteration, meaning the null in the visibilites is presented in the set of local minima. In addition, we note how the diversity of solutions increases. The red box encapsulates the most frequently repeated (most probable) solution, the blue box highlights the solution closest to the ideal, and the green box identifies the one with the lowest $\chi^2$.}
    \label{fig:paretos_86_allFalse}
\end{figure}

\begin{figure*}[h]
  \hspace{-2cm}
  \begin{subfigure}[t]{0.6\linewidth}
    \includegraphics[width=\linewidth]{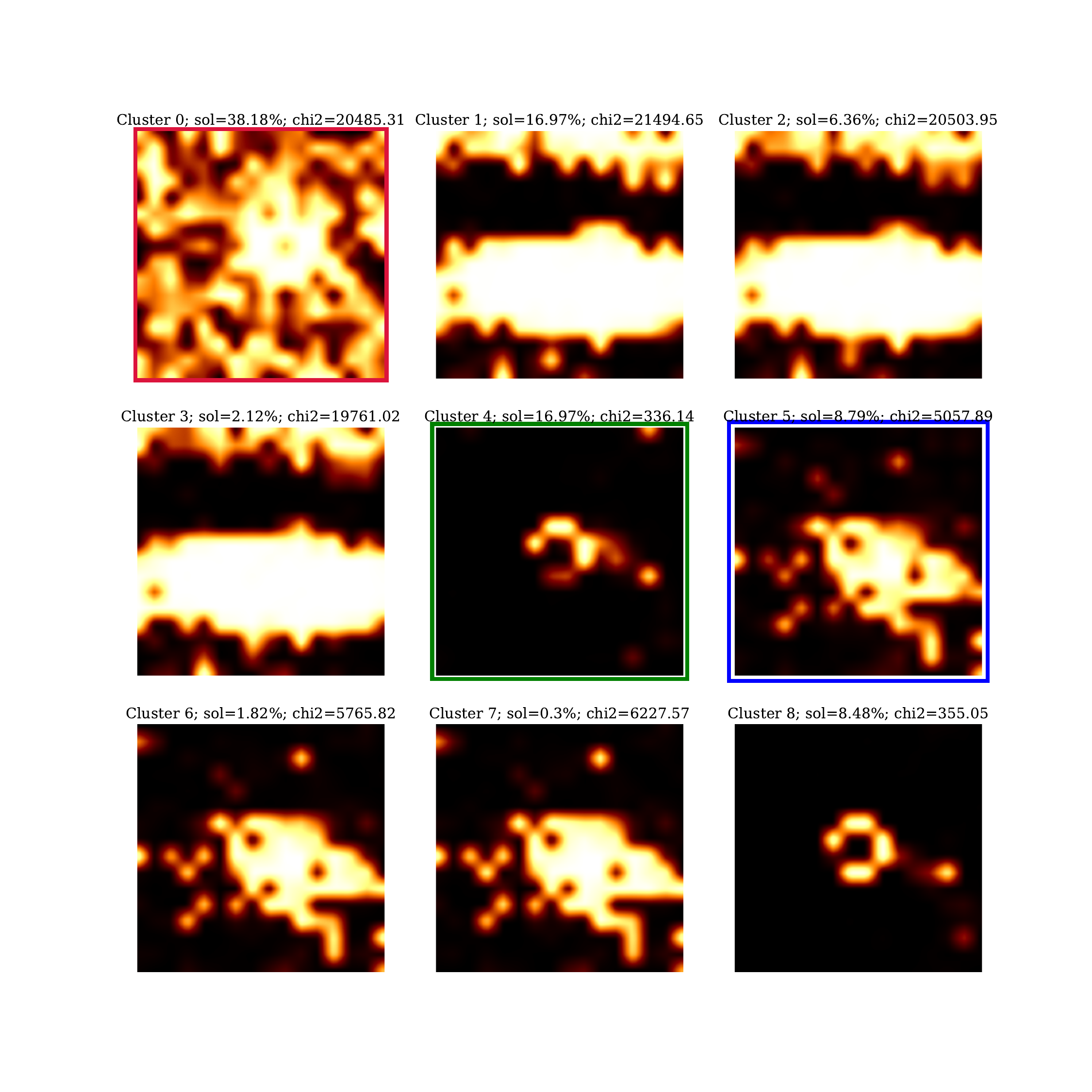}
    \vspace{-1cm}
    \subcaption{}  % (a)
    \label{fig:paretos_86_a}
  \end{subfigure}\hfill
  \begin{subfigure}[t]{0.6\linewidth}
    \vspace{-10cm}
    \includegraphics[width=\linewidth]{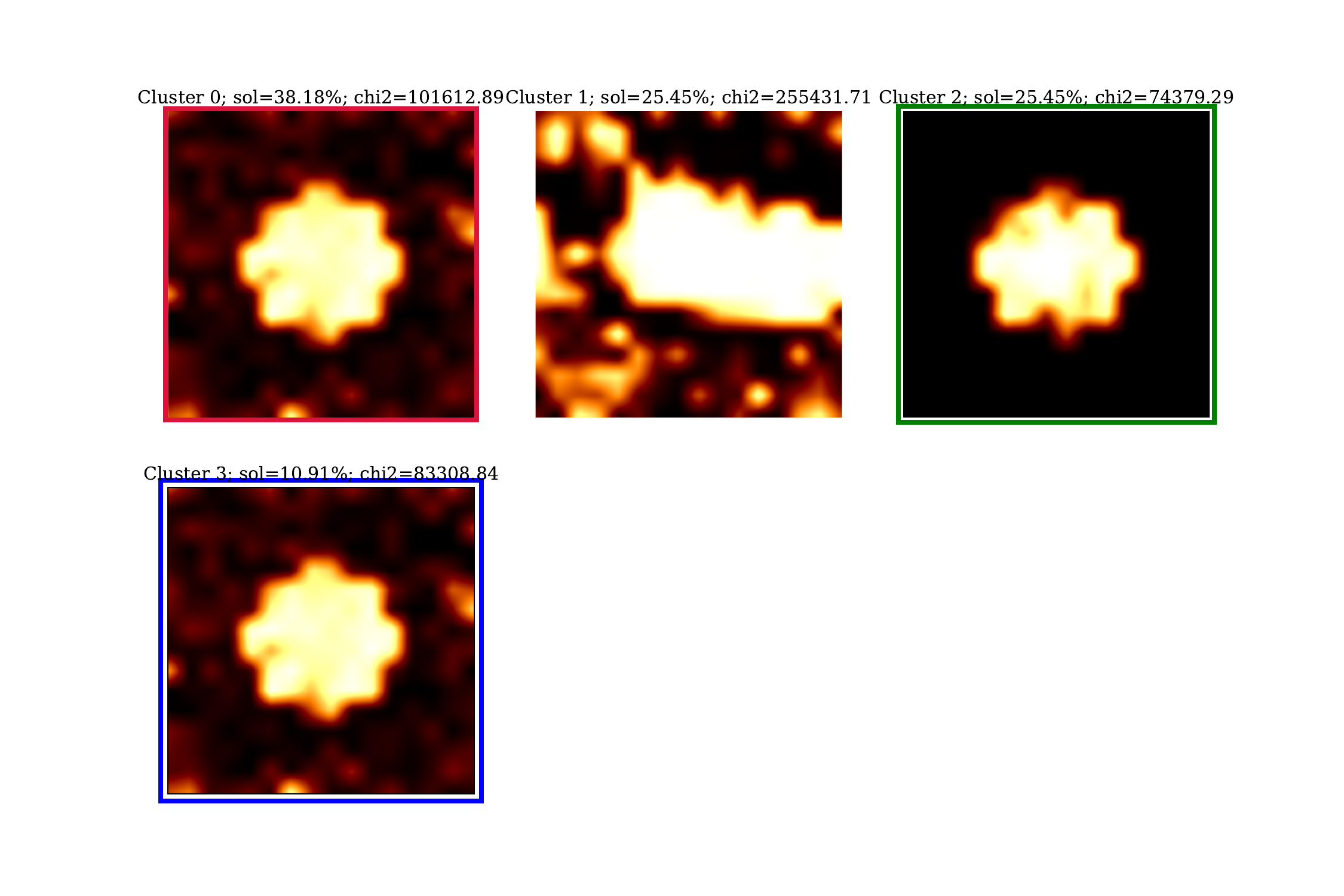}
    \vspace{-1cm}
    \subcaption{}  % (b)
    \label{fig:paretos_86_b}
  \end{subfigure}\hfill
  \begin{subfigure}[t]{0.6\linewidth}
    \centering
    \hspace{10cm}
    \includegraphics[width=\linewidth]{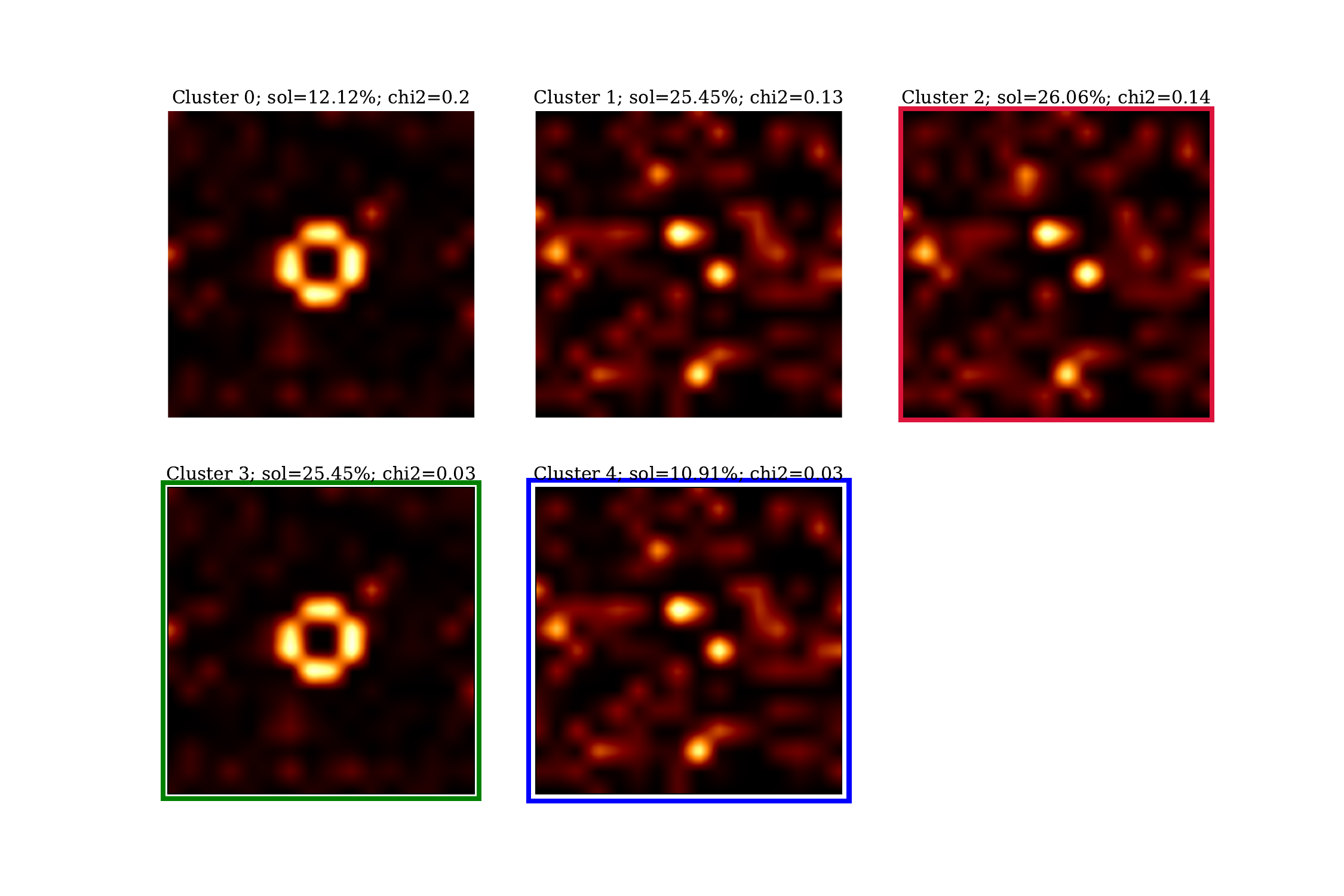}
    \vspace{-1cm}
    \subcaption{}  % (c)
    \label{fig:paretos_86_c}
  \end{subfigure}
  \caption{Pareto front for the intrinsic source at 86\,GHz.  (a) Gaussian~\cite[LSQ][]{Issaoun2019} starting point, (b) disk starting point, (c) ring starting point.  The red box encapsulates the most frequently repeated (most probable) solution, the blue box highlights the solution closest to the ideal, and the green box identifies the one with the lowest~$\chi^2$.}
  \label{fig:paretos_86}
\end{figure*}

\end{appendix}

\end{document}